\documentclass[12pt,preprint]{aastex}
\usepackage{natbib,graphicx,epstopdf}

\newcommand{\spitzer}{\emph{Warm-Spitzer}}
\newcommand{\kepler}{\emph{Kepler}}
\newcommand{\corot}{\emph{CoRoT}}

\newcommand{\gcmc}{\ensuremath{\rm g\,cm^{-3}}}
\newcommand{\rhk}{\ensuremath{R^{\prime}_{\rm HK}}}	
\newcommand{\logrhk}{\ensuremath{\log\rhk}}		
\newcommand{\sval}{\ensuremath{S_{\mbox{\scriptsize HK}}}}
\newcommand{\caii}{\ion{Ca}{2} H \& K}
\newcommand{\chisq}{\mbox{$\chi^2$}\,}

\newcommand{\teff}{\ensuremath{T_{\rm eff}}}

\newcommand{\rsun}{\ensuremath{R_\sun}}
\newcommand{\msun}{\ensuremath{M_\sun}}

\newcommand{\rstar}{\ensuremath{R_\star}}
\newcommand{\mstar}{\ensuremath{M_\star}}

\newcommand{\rpl}{\ensuremath{R_{\rm P}}}
\newcommand{\mpl}{\ensuremath{M_{\rm P}}}

\newcommand{\rhopl}{\ensuremath{\rho_{\rm P}}}

\newcommand{\teq}{\ensuremath{T_{\rm eq}}}

\newcommand{\rjup}{\ensuremath{R_{\rm J}}}
\newcommand{\mjup}{\ensuremath{M_{\rm J}}}

\newcommand{\kc}{Kepler-5}
\newcommand{\kcb}{Kepler-5b}
\newcommand{\ks}{Kepler-6}
\newcommand{\ksb}{Kepler-6b}

\newcommand{\hda}{HD~209458b}

\shorttitle{Kepler-5b and \ksb\ secondary eclipse measurements}
\shortauthors{Desert et al.}\def\simgr{\,\hbox{\hbox{$ > $}\kern -0.8em \lower 1.0ex\hbox{$\sim$}}\,}
\def\simle{\,\hbox{\hbox{$ < $}\kern -0.8em \lower 1.0ex\hbox{$\sim$}}\,}

\begin{document}

\title{The atmospheres of the hot-Jupiters \kcb\ and \ksb\ observed during occultations with \textit{Warm-Spitzer} and \textit{Kepler}}


\author{Jean-Michel D\'esert\altaffilmark{1},
David Charbonneau\altaffilmark{1},
Jonathan J. Fortney\altaffilmark{2},
Nikku Madhusudhan\altaffilmark{3},
Heather A. Knutson\altaffilmark{4},
Fran\c{c}ois Fressin\altaffilmark{1},
Drake Deming\altaffilmark{5},
William J. Borucki\altaffilmark{6},
Timothy M. Brown\altaffilmark{7},
Douglas Caldwell,\altaffilmark{8},
Eric B. Ford\altaffilmark{9},
Ronald L. Gilliland\altaffilmark{10},
David W. Latham\altaffilmark{1},
Geoffrey W. Marcy\altaffilmark{11},
Sara Seager\altaffilmark{12},
and the Kepler Science Team
}
\altaffiltext{1}{Harvard-Smithsonian Center for Astrophysics, 60 Garden Street, Cambridge, MA 02138; jdesert@cfa.harvard.edu}
\altaffiltext{2}{Department of Astronomy and Astrophysics, University of California, Santa Cruz, CA 95064, USA}
\altaffiltext{3}{Department of Astrophysical Sciences, Princeton University, Princeton, NJ 08544, USA}
\altaffiltext{4}{Department of Astronomy, University of California, Berkeley, CA 94720-3411, USA} 
\altaffiltext{5}{Solar System Exploration Division, NASA Goddard Space Flight Center, Greenbelt, MD 20771, USA}
\altaffiltext{6}{NASA Ames Research Center, Moffett Field, CA 94035}
\altaffiltext{7}{Las Cumbres Observatory Global Telescope, Goleta, CA 93117}
\altaffiltext{8}{SETI Institute, Mountain View, CA 94043}
\altaffiltext{9}{University of Florida, Gainesville, FL 32611}
\altaffiltext{10}{Space Telescope Science Institute, Baltimore, MD 21218}
\altaffiltext{11}{University of California, Berkeley, CA 94720}
\altaffiltext{12}{Massachusetts Institute of Technology, Cambridge, MA 02159, USA}

\begin{abstract}

This paper reports the detection and the measurements of occultations of the two transiting hot giant exoplanets \kcb\ and \ksb\ by their parent stars. The observations are obtained in the near infrared with \spitzer\ Space Telescope and at optical wavelengths by combining more than a year of {\it Kepler} photometry. 
The investigation consists of constraining the eccentricities of these systems and of obtaining broad band emergent spectra for individual planets. 
For both targets, the occultations are detected at $3~\sigma$ level at each wavelength with mid-occultation times consistent with circular orbits.
The brightness temperatures of these planets are deduced from the infrared observations and reach T$_{\rm{Spitzer}}$=$1930\pm100$~K and T$_{\rm{Spitzer}}$=$1660\pm120$~K for \kcb\ and \ksb\ respectively. 
We measure optical geometric albedos $A_g$ in the \kepler\ bandpass and find $A_g = 0.12 \pm 0.04$ for \kcb\ and $A_g = 0.11 \pm 0.04$ for \ksb\ leading to upper an limit for the Bond albedo  $A_B \leq 0.17$ in both cases.
The observations for both planets are best described by models for which most of the incident energy is redistributed on the dayside, with only less than $10\%$ of the absorbed stellar flux redistributed to the night side of these planets.
The data for \kcb\ favor a model without a temperature inversion, whereas for \ksb\ they do not allow distinguishing between models with and without inversion.

\end{abstract}


\section{Introduction}\label{intro}

The highly irradiated transiting hot-Jupiters currently provide the best opportunities for studying exoplanetary atmospheres in emission, during planetary occultations, when the exoplanets pass behind their parent stars. Emission from transiting hot-Jupiters (HJs) have been studied theoretically prior to any detection of light from an alien world \citep{seager00,sudarsky00}. The first exoplanetary emissions were first detected from space in the infrared wavelengths \citep{charbonneau05,deming05} and more recently in the optical bandpass \citep{alonso09a,snellen09,borucki09,alonso10} using the \corot\ space telescope. 
Notably, \cite{rowe06,rowe08} use the {\it Microvariablity and Oscillations of Stars} ({\it MOST}) telescope to place a very stringent upper limit on the optical depth of the occultation of \hda.
Recently, ground based spectrophotometry has also permitted observations of HJs in the near infrared (e.g. \citealt{sing09,lopez10,gillon09,croll10}).
The planet-to-star flux ratio is in the order of ten times higher at infrared wavelengths, due to the thermal emission from the planet, compared to optical wavelengths, domain where the starlight is reflected by the planet \citep{sudarsky03,fortney05,burrows05,seager05,barman05}. Consequently, only a few objects can be detected in the visible, those for which we will have enough observations to combine during occultations to improve the signal-to-noise and reveal the eclipse event. However, these objects are very valuable for understanding the energy budget of HJs and, to some extent, for comparative exoplanetology. The HJs detected by \corot\ and \kepler\ are particularly interesting since their occultations can be observed in the optical bandpass  \citep{snellen09,borucki09}.


Obtaining multiple wavelength measurements of  the relative depths of planetary occultations is fundamental to understanding the energy budget of these objects \citep{sudarsky03,burrows05}. The thermal structure of HJ is likely to be influenced by strong stellar irradiation. For example, high altitude optical absorbers can lead to atmospheric temperature inversions \citep{hubeny03,fortney08,burrows07,burrows08} as observed to a wide range of irradiation levels (e.g. \citealt{harrington07,knutson08,machalek08,todorov10}). Although the nature of the absorber is subject to debate, titanium oxide (TiO) and vanadium oxide (VO) are amongst the best molecular candidates, since they are expected to be present in these hot atmospheres and that they are strong absorbents of the incident visible stellar flux \citep{fortney08}.
Furthermore, non-equilibrium chemistry, due to photochemistry or vertical mixing, could potentially lead to strong UV or optical absorbers being found in the upper atmosphere of these planets. This can also play a role in inverting the thermal structure \citep{burrows08,zahnle09}, although degeneracies between thermal inversions and chemical composition are known to exist in atmospheric models \citep{madhu09}. Interestingly, the presence of high altitude absorbers in HJ atmospheres can potentially be revealed with transmission spectroscopy \citep{desert08}. Emission spectroscopy observations help also to probe the atmosphere of HJs \citep{knutson08,machalek09}.  \cite{spiegel10a} and \cite{spiegel10b} demonstrate how the \kepler-band optical flux from a hot exoplanet depends on the strength of a possible extra optical absorber in the upper atmosphere.
Therefore, multi-wavelength observations are necessary to determine the energy budget of these exoplanets.
Measurements of the optical and near infrared emergent fluxes from the planet allow us to derive the Bond albedo of these objects, to learn about the energy distribution and to address whether or not an atmospheric thermal inversion can be revealed. 
In the case of \hda, the {\it MOST} observations show that the planet has a very low albedos at optical wavelengths \citep{rowe08}, which rules out the presence of bright reflective clouds in this exoplanet's atmosphere.
In a similar framework, \cite{deming10} combine occultations from {\it CoRoT} and from various space and ground-based infrared measurements of CoRoT-1 and 2 \citep{gillon09,gillon10}, to probe the atmospheric structures of these HJs.  
The multi-wavelength observations show that CoRoT-1's spectrum is well-reproduced by a $2460~K$ black-body to first order, which is interpreted to be due to either a high altitude layer that strongly absorbs stellar radiance, or an isothermal region in the planetary atmosphere and that CoRoT-2 spectrum exhibits line emission from CO at 4.5 microns which could be caused by tidal-induced mass loss. 
The strength of the multi-wavelength approach is that it allows also to measure or place meaningful upper limits on the geometric albedo of a small set of transiting extrasolar planets. For example, the occultations of HAT-P-7b measured with the {\it EPOXI}, {\it Kepler} and {\it Spitzer} spacecrafts generated a broadband spectrum covering $0.35-8~\mu$m. They have also been used to identify a set of atmospheric models that reproduce the observations \citep{christiansen10}. The interpretation of this spectrum shows that it is consistent with an inefficient day-night redistribution as seen from the \kepler\ phase curve \citep{borucki09,spiegel10b}. However, the \kepler\ phase curve of this planet reveals a high night-side temperature at around $2600~K$ \citep{welsh10}, which appears to be at odds with knowing inefficient redistribution scenario {\it a priori}.

In this paper, we combine occultation measurements obtained in the optical with \kepler\ and in the infrared with \spitzer\ to learn about the atmospheric properties of two HJs, \kcb\ \citep{koch10b} and \ksb\ \citep{dunham10}, discovered using \kepler\ spacecraft. A search for occultations events was initially done using the two first Kepler quarters secured in long-cadence mode \citep{kipping10}. In the present study, the photometry of the host stars was mearly continuously monitored by the \kepler\ space telescope allowing us to gather more than a year of observations and combine multiple occultations to improve the signal-to-noise. These two planets have masses and radii which are common amongst the known transiting HJs \citep{latham10}, as shown in Table~\ref{tab:targets}. Therefore, they may be representative of their class which make them good candidates to study, especially if we want to generalize our conclusions for comparative exoplanetology. We also obtained measurements of these targets in the near infrared with the \spitzer\ space telescope during occultations. 
Because of the high temperature of these HJs, our optical measurements are mainly due to thermal radiation rather than purely reflected light.

We first describe the observations, time series and analysis for \spitzer\ data in Sect.~2 and \kepler\ data in Sect.~3 and finally discuss our results in Sect.~4.

%
%
%
%
%

\section{\spitzer\ observations and photometry}
\label{sec:spitzer}

This section describes the \spitzer\ observations and provides details of our extraction of the time series spanning each individual occultation event.

\subsection{Observations}

As \emph{Spitzer} exhausted its cryogen of liquid coolant on 15 May 2009, only the first two channels, at 3.6 (channel~1) and 4.5~\micron\ (channel~2), of the Infrared Array Camera (IRAC; \cite{fazio04}) are available in the post-cryogenic mission.
\kc\ and \ks\ were both visited twice at each available IRAC bandpass, leading to a total 8 eclipses being secured as part of the 800~hours allocated to the program PID~60028 (PI: D. Charbonneau). We present here the first observations from this program which contribute to characterizing and vetting \kepler\ candidates.
Each visit is secured in full frame mode ($256\times256$) with an exposure time of 10.4 seconds, at 12 second cadence, leading to 2700 frames per \kc\ (K=11.77) 10~hours observations and 2150 frames per \ks\ (K=11.71) 8~hours observations. The full set of \spitzer\ observations is presented in Table~\ref{tab:spitzer_obs}.

\subsection{Photometry}

We use the Basic Calibrated Data (BCD) frames produced by the standard IRAC calibration pipeline for the photometric extraction. These files are corrected for dark current, flat-fielding, and detector non-linearity and converted into flux units.  

The first step of the photometric extraction consists of determining the centroid position of the stellar point spread function (PSF) using DAOPHOT-type Photometry Procedures, \texttt{CNTRD}, from the IDL Astronomy Library \footnote{{\tt http://idlastro.gsfc.nasa.gov/homepage.html}}. We use the \texttt{APER} routine to perform an aperture photometry with a circular aperture of variable radius, using radii of $1.5$ to $6$ pixels, in $0.5$ steps. Finally, the propagated uncertainties are derived as a function of the aperture radius, and we adopt the aperture which provides the smallest errors. We find that the eclipse depths and errors vary only weakly with the aperture radius for all the light-curves analyzed in this project. In the case of \kc, the optimal apertures are around $5$ and $2.5$~pixels at 3.6 and 4.5~\micron\ respectively. \ks\ has a companion at an angular separation of $4.1\arcsec$ and which is 3.8 magnitudes fainter \citep{dunham10}. This star is located at 3.5 IRAC pixels from our target of interest. Therefore, we fix our circular aperture at a radius of 2.5 pixels for all the visits of this star in order to minimize the contribution of the companion. 

The background level for each frame is determined by two methods.
We use \texttt{APER} to measure the median value of the pixels inside an annulus
centered on the star with inner and outer radii of $9$ and $16$
pixels, respectively. 
We also estimate the background by fitting a Gaussian to the central region of the histogram of counts from the full array. The center of the Gaussian fit is adopted as the residual background intensity.
These two methods provide similar values; therefore, we use the background vlaues determined from an annulus.

The contribution of the background to the total flux from the stars \kc\ and \ks\ 
is low in both IRAC band-passes, from 0.2\% to 0.55\% depending of the photometric aperture size. Therefore, photometric errors are not
dominated by fluctuations in the background. 
We note that the background is negative for all channel~1 observations and for one observation secured in channel~2 .
The 10.4 seconds exposures yield a typical signal-to-noise ratio ($S/N$) of $200$ and $150$ per individual observation at 3.6 and 4.5~\micron\ respectively.

After producing the photometric time series, we use a sliding median filter that compares the 20 preceding and 20 following photometric measurements and centroid positioning to identify and trim outliers greater than $5~\sigma$. This process removes measurements affected by transient hot pixels and inaccurate centroid determination.  In this way, we discarded approximately $2\%$ photometric points from all the observations.  We also discarded the first half hour of observations, corresponding to around a hundred frames.
These frames exhibit an anomalously large pointing drift, most likely due to settling of the telescope at the new position.
The final number of photometric measurements used for each observation is given in Table~\ref{tab:spitzer_obs}.

\subsection{Occultation amplitudes and associated errors}
\label{model_eclipse}

To measure the occultation depths and their uncertainties we model the light curves with 4 parameters: the occultation depth, $d$, the orbital semi-major axis to stellar radius ratio (system scale), $a / R_\star$, the
impact parameter, $b$, and the time of mid transit, $T_c$.  We
use the IDL transit routine \texttt{OCCULTSMALL}, developed by
Mandel \& Agol (2002), to model the light curve. 
\kcb\ and \ksb\ have both well defined transit and stellar parameters constrained by the transits from the \kepler\ light-curves (Koch et al. 2010, Dunham et al. 2010). Thus, we adopt these parameters, in particular the inclination and the scale of the system, by fixing $a / R_\star$ and $b$ to their nominal values in our eclipsing model. We also fixed the mid-eclipse times to the values we derived from the \kepler\ lightcurves (see Sect.~\ref{sec:kepler} below). Only the depth of the occultations are allowed to vary when fitting the \spitzer\ observations.


The \spitzer/IRAC photometry is known to be systematically affected by
the so-called \textit{pixel-phase effect}, which is due to the combination of pointing jitter and intra-pixel sensitivity (see, e.g., Charbonneau et al. 2005, 2008,
Reach et al. 2006). This effect corresponds to oscillations in the
measured raw light curve with an approximate period of $70$~minutes and an amplitude of $2\%$ peak-to-peak. This artefact has to be corrected to properly extract the eclipse depths and errors.
To correct the raw light curve for this intra-pixel sensitivity effect, we use the centroid position of the target on the detector and its variations as function of time. We use a quadratic function of the position and a linear function of time 
$F_{\mathrm{corr}}=F[K_1(x-x_0)+K_2(x-x_0)^2+K_3(y-y_0)+K_4(y-y_0)^2+K_5+K_6\times t]$,
where $F$ and $F_{\mathrm{corr}}$ are the fluxes of the star before and after the pixel-phase 
effect correction, and $(x-x_0)$ and $(y-y_0)$ define the position in pixel of the source centroid on the detector with respect to the pixel pointing position, located at $[x_0,y_0]$, and the constants $K_i$ are the free parameters to adjust.

We use the \texttt{MPFIT} package\footnote{{\tt http://cow.physics.wisc.edu/$\sim$craigm/idl/idl.html}} to perform
a Levenberg-Marquardt least-squares fit of the transit model to observations.  The best-fit model is computed over the whole parameter space ($d$, $K_i$).  The baseline function described above is combined with the transit light curve function
so that the fit is constrained by $8$ parameters (2 for the transit model, 2 for the linear baseline, and 4 for the pixel phase effect).

We use four methods to determine the centroid position of a stellar PSF and test the its impact on the final results.
In the first method, we fit Gaussians to the marginal x and y sums using \texttt{GCNTRD}.
The method we apply consists of computing the centroid of the stars using a derivative search with \texttt{CNTRD} also part of the standard IDL Astronomy Library. 
As for the third method, we calculate the center-of-light of the star within an circular aperture with a radius of $3.0$ pixels to approximate the center of the star.
Finally, the fourth method uses a symmetric 2D Gaussian fit with a fixed width to a 3x3 pixel sub-array to approximate the center of the star, as suggested by \citet{agol10}.

All these methods allow us to measure the centroid position of a stellar image with a different level of accuracy and precision as previously noted by \citet{agol10}. 
For all the methods tested, we find that the centroid position varies by less than $20\%$ of a pixel during a complete observation and that can be determined to a precision of a hundredth of a pixel. 
We measure the eclipse depths and associated errors for all the four methods following the algorithm we describe below.
Although each method provides slightly different results for the centroid determination, it yields an eclipse depth consistent to within $1~\sigma$ and similar errors. 
In this paper we use the \texttt{CNTRD} program as it produces the smallest reduced \chisq.

As shown by Pont et al. (2006), the existence of low-frequency correlated noise (red noise) between different exposures must be considered to obtain a realistic estimation of the uncertainties. To obtain an estimate of the systematic errors in our observations we use the permutation of the residuals method, known as ``prayer-bead'' (Moutou et al. 2004, Gillon et al. 2006, 2007) and derived the covariance from the residuals of the light curve. 
In this method, the residuals of the initial fit are shifted systematically and sequentially by one frame, and then
added to the transit light curve model before fitting again. The error on each photometric point is the same and is set to the
$rms$ of the residuals of the first best-fit obtained. We find that the final $rms$ values are $5$ to $20\%$ larger than the predicted photon noise depending on the target and the bandpass.

We produce as many shifts and fits of transit light curves as the number of photometric measurements to determine the statistical and systematical errors for the adjusted parameters. 
We set our uncertainties equal to the range of values containing $68\%$ of the points in the distribution in a symmetric range about the median for a given parameter. We check that these values are close to the standard deviation of each nearly Gaussian distributions. We fit for the eclipse depth fixing the mid-eclipses time to the values we derived from the \kepler\ observations as describe in the following section. The best-fits of the occultation depths their error bars are listed in Table~\ref{tab:spitzer_obs}.

\section{Analysis of the \kepler\ light curves}
\label{sec:kepler}

This section describes the \kepler\ observations gathered almost continuously during nearly 14 months and provides details on the analysis of the different occultations.

The \kepler\ pass-band spans $437$ to $897~nm$, with a central wavelength roughly equivalent to the R-band \citep{koch10a,batalha10,bryson10}.
The stars \kc\ and \ks\ have \kepler\ magnitudes ($Kp$) of 13.37 and 13.30 \citep{koch10b,dunham10}. The light of the companion star close-by \ks\ as discussed above in Sect.~\ref{sec:spitzer} is included in the \kepler\ aperture. This has the effect of diluting the depth of the eclipse signal by only a few percent as shown by \citet{dunham10}, well below the precision we obtain (see below).

We use the \kepler\ science data of \kc\ and \ks\ from Quarter~0 to 5 (Q0-Q5). These observations have been reduced and detrended by the \kepler\ pipeline \citep{jenkins10a}. 
The pipeline produces both calibrated light curves (PA data) for individual analysis and corrected light curves (PDC) which are used to search for transits. We used both datasets and checked that the conclusions remain the same.
This paper presents results which are measured from PDC data.
They consist of long cadence integration time (30~minutes) for Quarters~0 and 1 \cite{caldwell10,jenkins10b} and short cadence (1~minute) for Quarters~2 to 4 \cite{gilliland10}. 
We acquired the data in a uniform manner at short cadence between BJD 2454998 and 2455371 (UT 2009 June 15 - 2010 June 23) during 373 days of observations (Q2-Q5). The pipeline provides time series with times in barycentric corrected Julian days, and flux in electrons. 

\begin{figure}[h!]
\begin{center}
\includegraphics[width=3in]{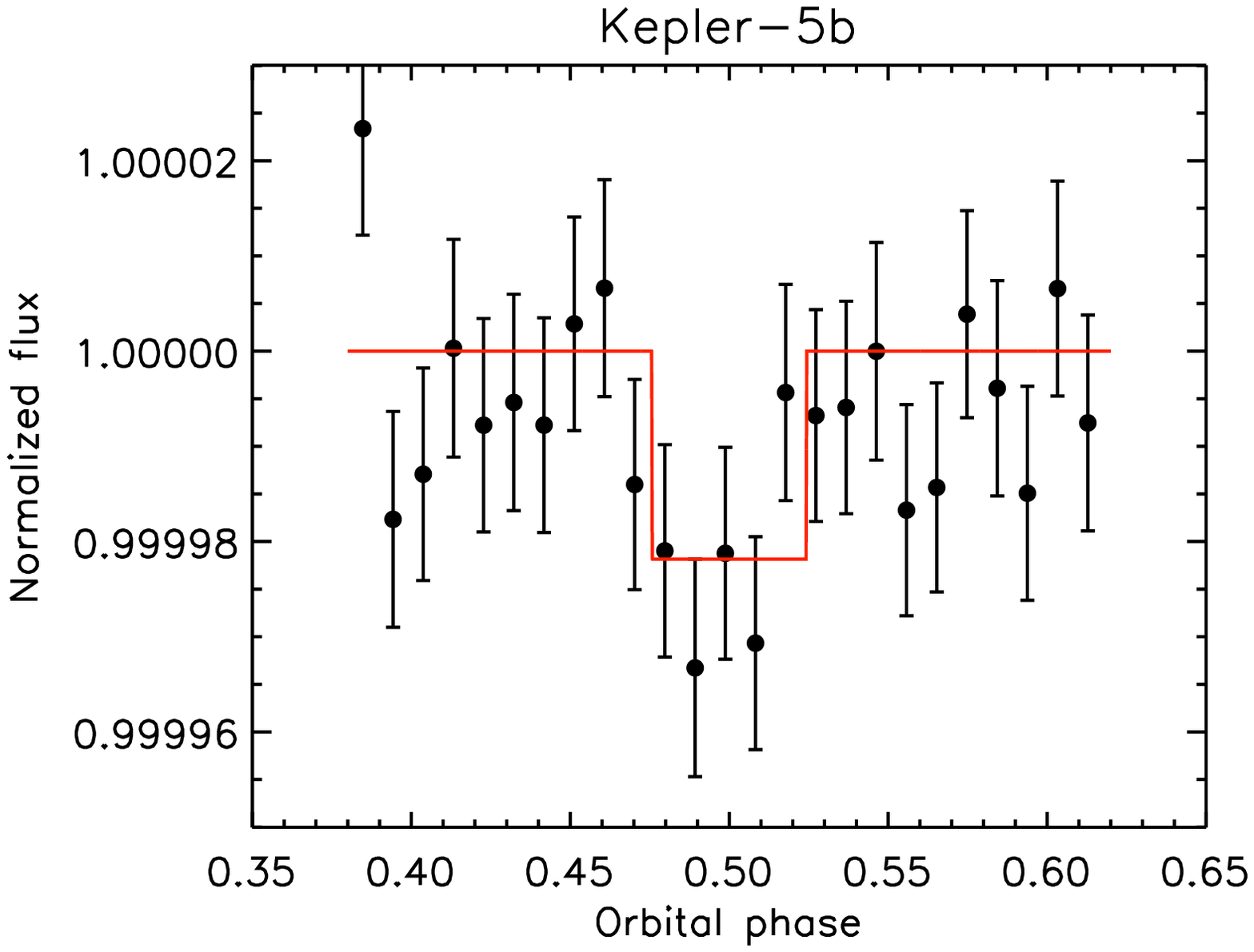}
\includegraphics[width=3in]{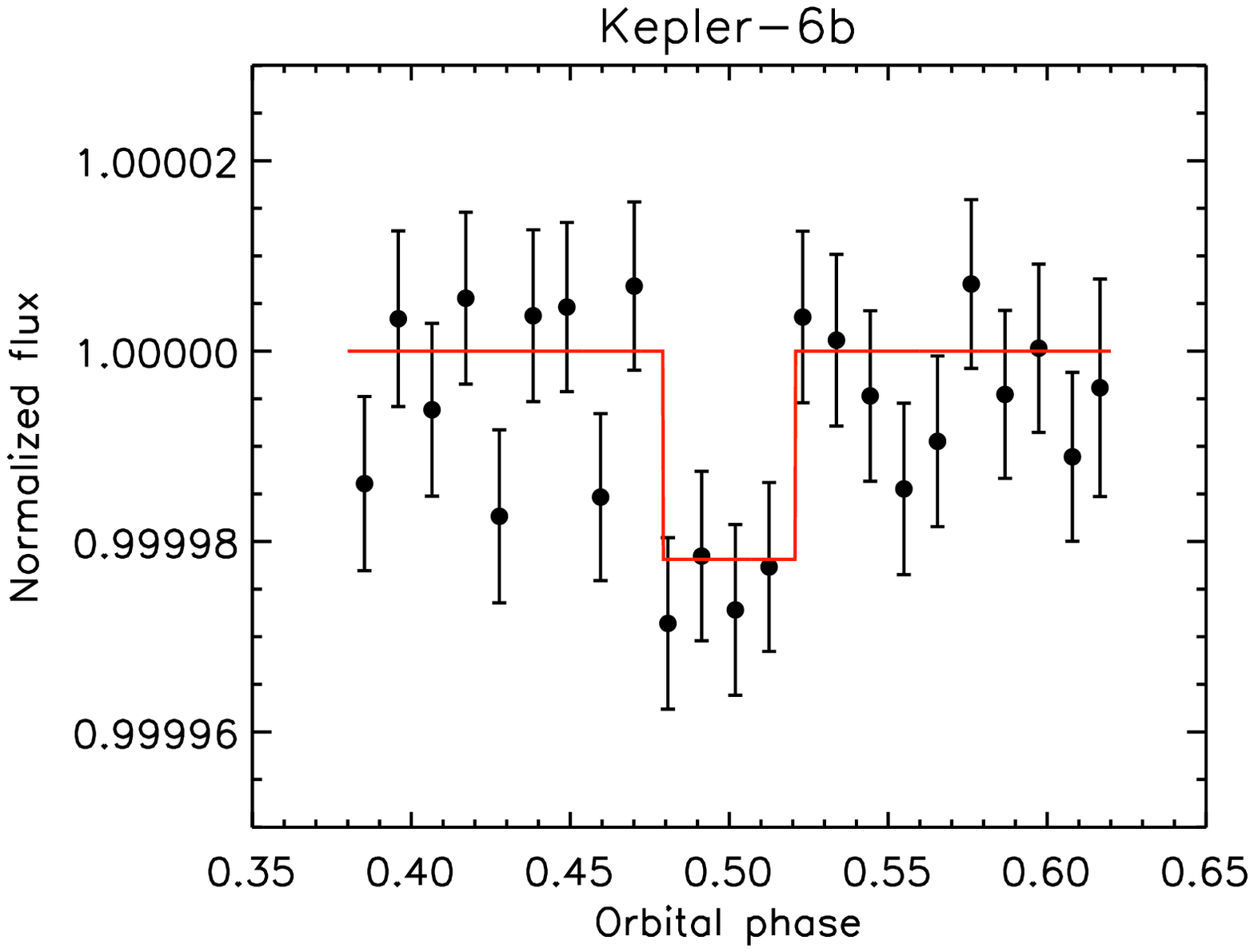}
 \caption{Phase folded normalized \kepler\ occultation light-curves. Plot for \kcb\ is on the left side and \ksb\ is on the right side. The observation are the black point with their one-sigma error-bars. The data are binned by approximately 40~minutes (0.01 in phase). The best fit are over-plotted with the red solid line.}
   \label{fig:foldlightcurves}
\end{center}
\end{figure}

The quality of the photometric time series around each expected eclipse event is checked visually in a first step. This step allows to flag and remove eclipses that we consider not satisfactory for various reasons (artefacts, Spacecraft-related events occasionally result in loss and subsequent reacquisition of fine guidance, etc...). We reject $20$ of the $103$ occultation events observed for \kc, and $12$ of these events being amongst $113$ for \ks. The numbers of occultations that we keep per target and per observing mode are presented in Table~\ref{tab:kepler}.  
We then identify and trim outliers greater than $5~\sigma$ using a sliding median filter that compares the $20$ preceding and $20$ following photometric measurements. This process rejects a very small number of photometric measurements (less than 0.01\%) compared to the total measurement number (102,300 for \kc\ and 114,000 for \ks).
At this point we check that the $rms$ scatter is constant over the whole observational period.
We fit each individual eclipses between the planetary orbital phase $0.38$ to $0.62$ using the occultation model (see description in Sect.~\ref{model_eclipse})  assuming constant ephemerides and null limb darkening coefficients. All the transit parameters but the occultation depths and the mid-occultation times are set to fixe values. A linear function of time represents the baseline. Each individual fitted light-curves is then normalized to its local baseline in order to produce a set of normalized eclipse light curves. 
We finally phase fold and combine all the normalize light curves to produce four photometric time series, one for each observing mode (cadence) and target. 
We present the normalized, folded and combined light curves obtained at short cadence for both targets in Figure~\ref{fig:foldlightcurves} from which we measure the occultation depths as described in the following paragraphs.

\begin{figure}[h!]
\begin{center}
\includegraphics[width=3in]{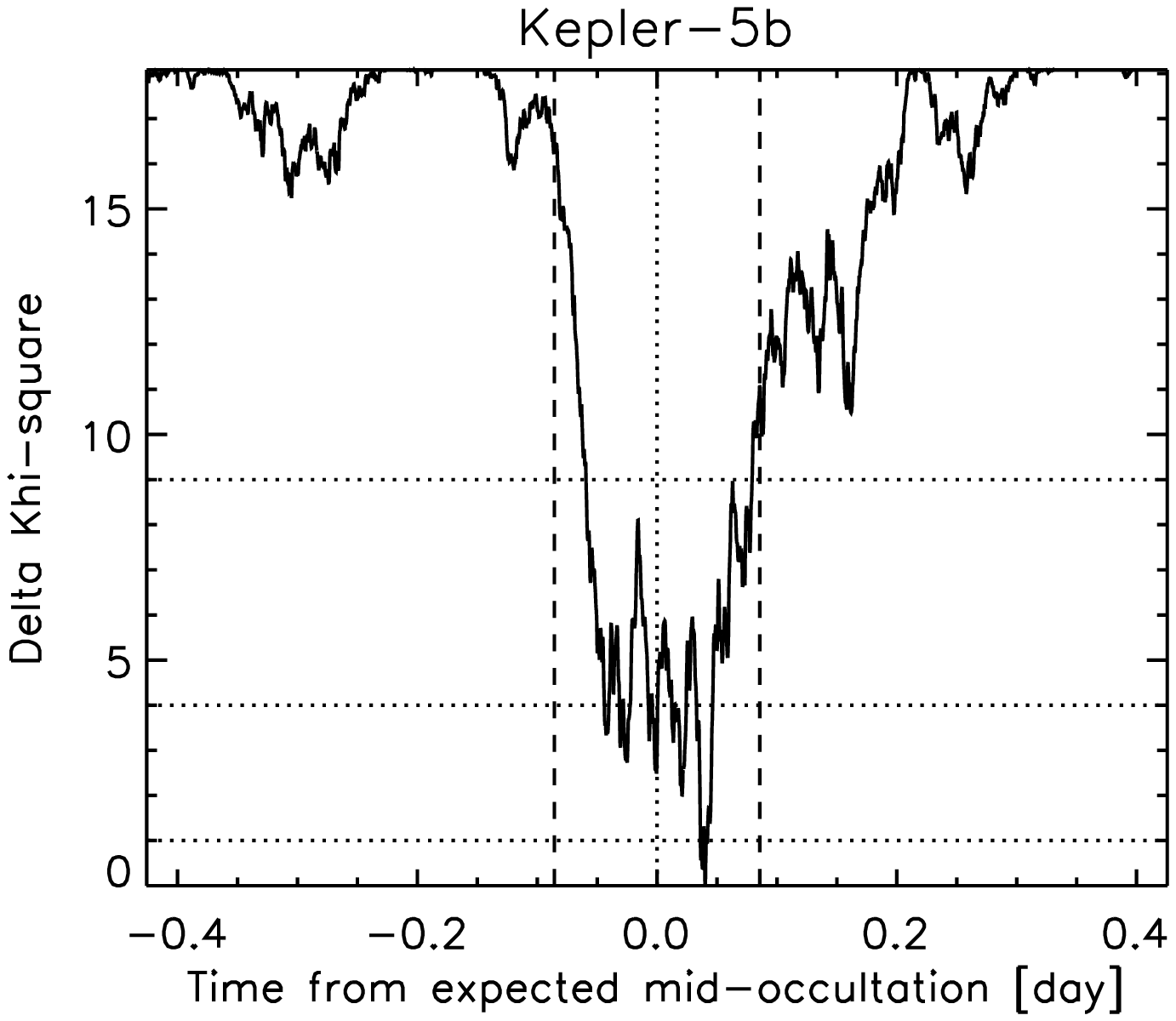}
\includegraphics[width=3in]{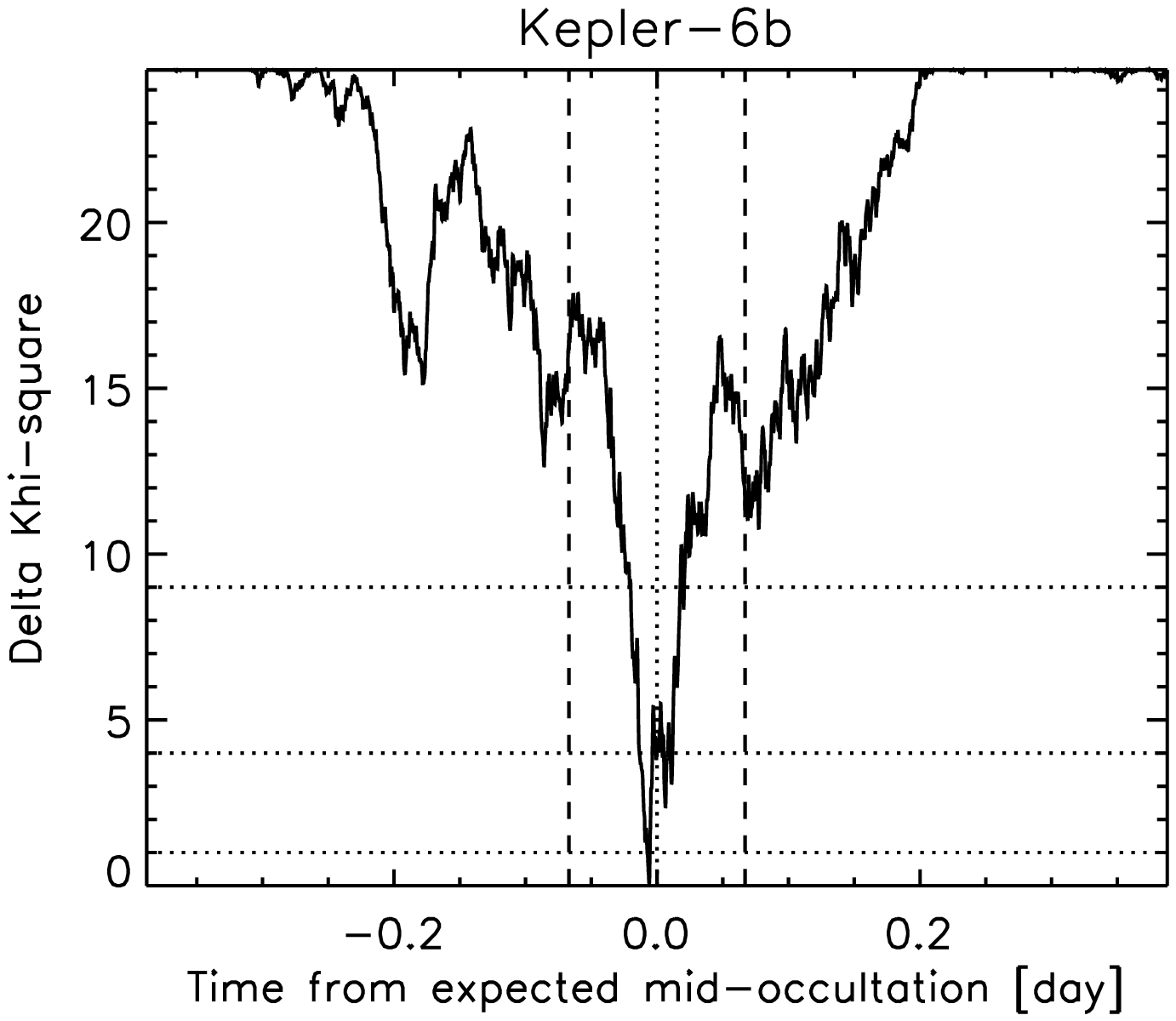}
 \caption{Delta \chisq\ from the best fit of the depth of a model as function of the mid-occultation time from zero (phase 0.5). Plot for \kcb\ is on the left side and \ksb\ is on the right side. The shape and duration of the occultation model are set by the transiting parameters. 
The vertical dashed lines show the transit durations and around zeros which are highlighted with vertical dotted lines.
Mid-occultations are expected at phase 0.5 for non eccentric orbits. The horizontal dotted lines indicates the limits of 1, 2 and $3~\sigma$ from the best fit.
The maximum occultation depth is centered near phase 0.5 which confirm the detection of the occultations event.}
   \label{fig:depthphase}
\end{center}
\end{figure}

We first search for the occultation events in the combined and normalized occultation light curves. This is done by evaluating the depths of Levenberg-Marquardt fits to models with the same overall shape as the planetary transits (same impact parameters and system scales), but at different phases of the orbital period. The best fits as function of the orbital phases are plotted in Fig.~\ref{fig:depthphase}, and the maximum depths and best fits are found very close to the orbital phase of $0.5$ as expected for a circular orbit. This suggests that we detect the occultations for both targets in the \kepler\ observations. 

\begin{figure}[h!]
\begin{center}
\includegraphics[width=3in]{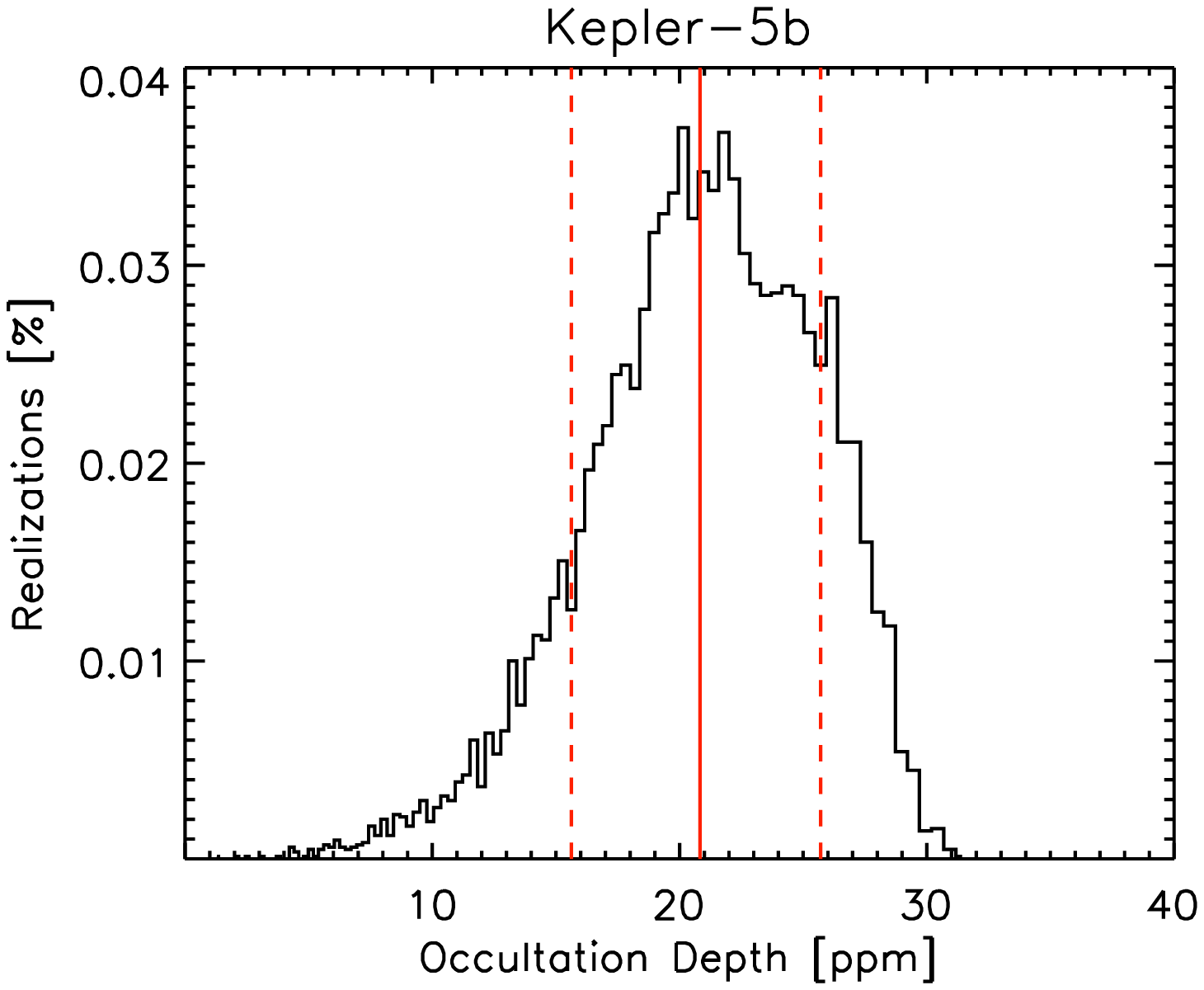}
\includegraphics[width=3in]{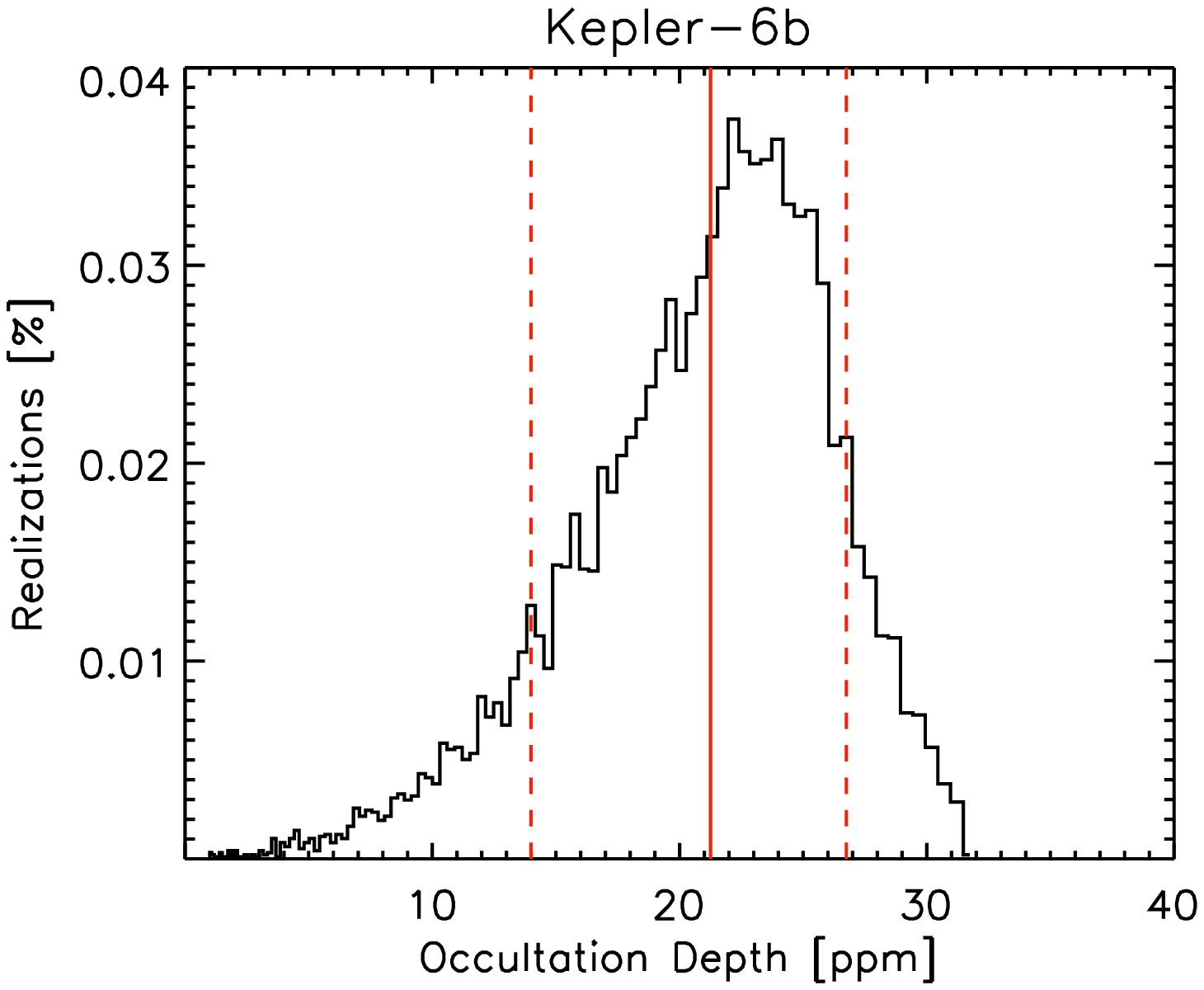}
 \caption{Measured distributions of the occultation depths obtained from the bootstrap trials. Distribution for \kcb\ is on the left side and \ksb\ is on the right side. Vertical red continuous lines correspond to the median of the distribution. Vertical red dashed lines
correspond to plus or minus 68\% of the distribution.}
   \label{fig:phasedepthchisq}
\end{center}
\end{figure}

We now estimate the  significance of these detections by measuring the occultations depths, ephemerides and associated errors for each observing cadence mode and targets.
We estimate the parameter values and errors using a bootstrap Monte-Carlo analysis. In this method, we first find the best-fit to the occultation curves and produce a set of photometric residuals. We compute the $rms$ of the residuals and create a new simulated data set by adding the errors, chosen randomly from a Gaussian distribution having the $rms$ for its FWHM, to the best-fit. We fit the new simulated data set and repeat this process for $10^5$ trials. We compute the median value and the standard deviation of the Gaussian distributions to derive the occultation depth and the central phase errors. The depths and mid-occultations as well as measurement significances are plotted as function of the orbital phases in Figure~\ref{fig:phasedepthchisq} and presented in Table~\ref{tab:kepler}. We weighted average the best-fitted depths and times to obtain one single value per target. We find that both planet have similar occultation depths of $21 \pm 6$ and $22 \pm 7$ ppm for \kcb\ and \ksb\ respectively.


\section{Results}\label{results}

We have shown that decreases in the fluxes of the \kepler\ and \spitzer\ light-curves correlated with the expected phases of occultations are detected in all band-passes at $3~\sigma$ level. We thus interpreted these signals as planetary occultations detected in the optical and in the near infrared wavelengths.

\subsection{Constraining the orbital eccentricity}
\label{ecc}

The current upper limit on the orbital eccentricity $e$ from radial velocity measurements is consistent with zero for both targets (Koch et al. 2010, Dunham et al. 2010). An independent analysis of the two first \kepler\ quarters (Q0 and Q1) have already been applied \citep{kipping10}. It uses the transits, eclipses and radial velocity measurements to confirm that no eccentricity is found for \kc\ and to report a marginal eccentricity at $2~\sigma$ level for \ks.

We measure the mid-occultation timing offset from both \kepler\ and \spitzer\ observations. The determination of the timing of the secondary eclipse constrains the planet's orbital eccentricity. A non-zero value of $e$ could produce a measurable shift in mid-eclipses times.
The \spitzer\ observations are affected by systematics (correlated noise) which prevent us from deriving timing measurements with a better precision than the precision we derive from \kepler\ measurements. Indeed, we measure the mid-occultation timing offset from \kepler\ with uncertainties of nearly 8 minutes for both targets.
Our estimate for the best-fit timing offset translates to a constraint on $e$ and the argument of pericenter $\omega$. The timing is used to constrain the $e\cos(\omega)$ at $3~\sigma$ level. For \kcb\ we find $e\cos(\omega)=-0.025 \pm 0.005$, and $|e\cos(\omega)|<0.06$ to $3~\sigma$. For \ksb\ we find $e\cos(\omega)=-0.01 \pm 0.005$, and $|e\cos(\omega)|<0.035$ to $3~\sigma$. 
These upper limits imply that the orbits of these objects are nearly circular unless the line of sight is aligned with the planet's major axes, i.e. the argument of periapse $\omega$ is close to 90\degr~or 270\degr.
Assuming that  both planets are on a circular orbit and accounting for the time taken by the light to cross the entire orbits, we would expect that the mid-secondary eclipses are delayed by $50$ seconds and $45$ seconds for \kc\ and \ks\ respectively, which translate into orbital phase of $0.50016$ for both targets.

\subsection{Atmospheric considerations}\label{atmo}

The occultation depths measured in each bandpass are combined and are turned into an emergent spectrum for each planet.
The observed flux of the planet in each bandpass corresponds to the sum of the reflected light, the thermal emission of the incident stellar flux, and the interior flux from the planet itself (e.g., internal heat or emission due to tidal forces). As a first step, we assume that the thermal emission from the planet itself is negligible.
We use the chromatic information to distinguish among the different components of the observed planetary flux.

If we were to explain the occultation detections in the \kepler\ bandpass as reflected light for both objects, it would imply that we are indeed measuring their geometric albedos. 
One can estimate the reflected light by the planet  as $\frac{F_p}{F_{\star}}= A_g\left(R_p\over a\right)^2$ where $A_g$ is the geometric albedo in the \kepler\ bandpass. With this assumption in mind, and with the occultation depths of  $21 \pm 6$ ppm  and  $22 \pm 7$ ppm for \kcb\ and \ksb\ respectively (see Table~\ref{tab:kepler}), we find that both planets have low geometric albedos in the \kepler\ band-pass. Thus, we obtain $A_g=0.12 \pm 0.04$ and  $A_g=0.11 \pm 0.04$ for  \kcb\ and \ksb\ respectively. Interestingly, we can infer the Bond-albedo $A_{\rm B}$ from $A_g$. 
The Bond albedo is the fraction of the bolometric incident radiation that is scattered back out into space at all phase angle.
Since we have no phase curves information for both objects, we assume a Lambertian criteria where $A_{\rm B} \leq 1.5 \times A_{\rm g}$, we which leads to $A_{\rm B} \leq 0.17$ at 1$~\sigma$ level for both object.

The equilibrium temperature is derived from
\begin{equation}
T_{eq}=T_\star(R_\star/a)^{1/2}[f(1-A_B)]^{1/4}
\label{eq} 
\end{equation}
which depends on the Bond albedo $A_B$ and the re-distribution factor $f$ which accounts for the efficiency of the transport of energy from the day to the night side of the planet. $f$ can vary between $1/4$ for an extremely efficient redistribution (isothermal emission at every location of the planet) and higher values for an inefficient redistribution, implying big differences in the temperatures between the day and the night sides of the planet.
Nevertheless, the detection of the occultations in the \kepler\ bandpass may indicate that we measure the thermal emission of these planets at these wavelengths, as expected for the temperature regime of HJs. Therefore, the Bond-albedo could be indeed well below 0.17.
Notably, as \kepler\ spacecraft continues to monitor the photometry of \kcb\ and \ksb, the observations will provide better constraints on occultations and may reveal the phase curves which would allow to fully constrain the Bond-albedo of these planets.

We estimate the thermal component of the planet's emission in the two \spitzer\ band-passes from the occultation depths measured at 3.6 and 4.5~\micron\ (see Table~\ref{tab:spitzer_obs}). To do so, we assume that the planetary emission is well reproduced by a black-body spectrum and translate the measured depth of the secondary eclipse into brightness temperatures. We use the PHOENIX atmospheric code \citep{hauschildt99} to produce theoretical stellar models for the star, 
with stellar temperature of $\teff = \ensuremath{6297\pm60}$ for \kcb\ (Koch et al. (2010)  $\teff = \ensuremath{5647\pm44}$ for \ksb\ (Dunham et al. 2010). Taking the \spitzer\ spectral response function into account, 
the ratio of areas of the star and the planet and the stellar spectra, we derive the brightness temperatures that best fit the observed eclipse depths measured in the two IRAC bandpasses. 
The brightness temperature calculated this way resulted in T$_{\rm{Spitzer}}$=$1930\pm100$~K and 
T$_{\rm{Spitzer}}$=$1660\pm120$~K for \kcb\ and \ksb\ respectively.  If we further assume that the planet is in thermal equilibrium and has a zero Bond albedo, these temperatures favor low values of the re-distribution factor $f$, i.e. the planet day-side re-radiate efficiently the incoming stellar energy flux. 

The brightness temperatures we derive using the \spitzer\ band-passes are larger than the equilibrium temperatures for both targets (see Table.~\ref{tab:stars}). This may indicates that our assumption that these planets behave as black-bodies is most probably wrong, similar to the giant planets of our solar system. Furthermore, a more robust determination of $A_{\rm B}$ requires detailed model computation because of the wide range of parameter space \citep{seager05}. We discuss below our results from two different types of atmospheric models.

%
%
%
%

In the first approach, we use the atmospheric retrieval method of \citet{madhu09} to derive the temperature structure and composition of each system, given the data. 
The model involves 1-D line-by-line radiative transfer, with parametric temperature structure and composition, and includes the major molecular and continuum opacity sources, along with constraints of LTE, hydrostatic equilibrium, and global energy balance. 
This modeling approach allows one to compute large ensembles of models ($\sim 10^6$), and explore  the parameter space of molecular compositions and temperature structure in search of the best-fitting models.
In the present work, the dominant sources of opacity included are: H$_2$O, CO, CH$_4$, CO$_2$, NH$_3$, and H$_2$-H$_2$ CIA absorptions, in the infrared, and TiO, VO, Na, K, and Rayleigh scattering, in the visible. In the present context, the number of model parameters is $N = 10 - 15$ \citep{madhu09}. Thus, the limited number of available observations ($N_{\rm obs} = 3$)  imply a substantial degeneracy in solutions. Consequently, our goal with the present data is to find a nominal set of solutions, as opposed to finding unique fits.  

Models fitting the observations of \kcb\ and \ksb\ are shown in Figure~\ref{fig:madhumodel}. We find that observations of \kcb\ can be explained to good precision by a wide range in composition, including models with equilibrium chemistry assuming solar abundances. However, current observations tentatively rule out a thermal inversion in this system. At the temperatures of \kcb\, the atmosphere is expected to be abundant in CO \citep{burrows99}, which has a strong feature in the 4.5 $\micron$ bandpass. A thermal inversion would therefore naturally predict a high flux in this channel over the 3.6 $\micron$ channel, as opposed to the observed fluxes which are similar between the two channels. Instead, the CO absorption feature caused by the lack of a thermal inversion explains the observations very well. Additionally, the \spitzer\ and \kepler\ data together require that the incident energy is mostly re-radiated on the dayside, with a low albedo and a small fraction ($f$) of dayside energy redistributed to the night side. Assuming zero albedo, the model shown has $f$ = 0.12. 

The observations of \ksb\ allow for the possibility of a thermal inversion in its atmosphere. As shown in Figure~\ref{fig:madhumodel}, the data can be explained to within the 1-$\sigma$ errors by models with and without thermal inversions. As explained above, the higher flux in the 4.5 $\micron$ channel over the 3.6 $\micron$ channel can be interpreted as a sign of an inversion. However, given the degrees of freedom allowed by the molecular abundances, the limited data can be fit just as well without inversions (see e.g. Madhusudhan \& Seager, 2010b). A wide range of abundances can fit the data, including those close (within a factor of ten) to  equilibrium chemistry assuming solar abundances. However, as with \kcb\, the \spitzer\ and \kepler\ data together place stringent constraints on the energy budget on the dayside, requiring low albedos and low redistribution to the night side.


The second model we consider to compare our data to is the HJ atmospheric model of \cite{fortney08}.
Our motivation here is that, given our limited wavelength coverage, it is important to compare our best-fit planetary-to-star flux ratio values with physically motivated models.
We aim at broadly distinguishing these atmospheres between different classes of models, given that there are no prior constraints on the basic composition and structure.

The atmospheric spectrum calculations are performed for 1D atmospheric pressure-temperature (P-T) profiles and use the equilibrium chemical abundances, at solar metallicity, described in \cite{lodders02, lodders06}. This is a self-consistent treatment of radiative transfer and chemical equilibrium of neutral species. The opacity database is described in \cite{freedman08}. So far, this model has been used to generate P-T profiles for a variety of close-in planets \cite{fortney05,fortney06,fortney08}.
The models are calculated for various dayside-to-nightside energy redistribution parameter ($f$) and allow for the presence of TiO at high altitude, playing the role of absorber which likely lead to an inversion of the T-P profile.

\begin{figure}[h!]
\begin{center}
\includegraphics[width=5in]{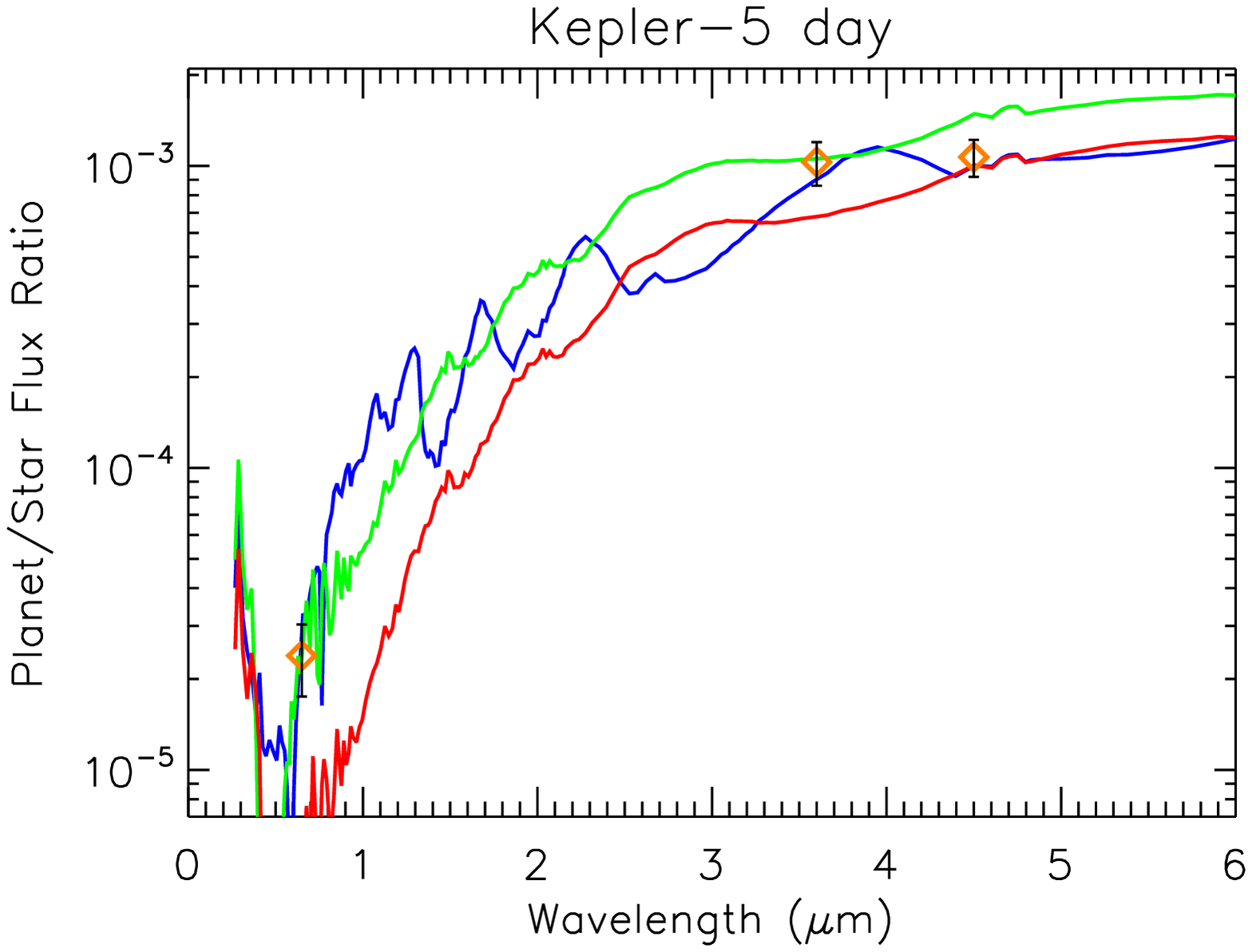}
\includegraphics[width=5in]{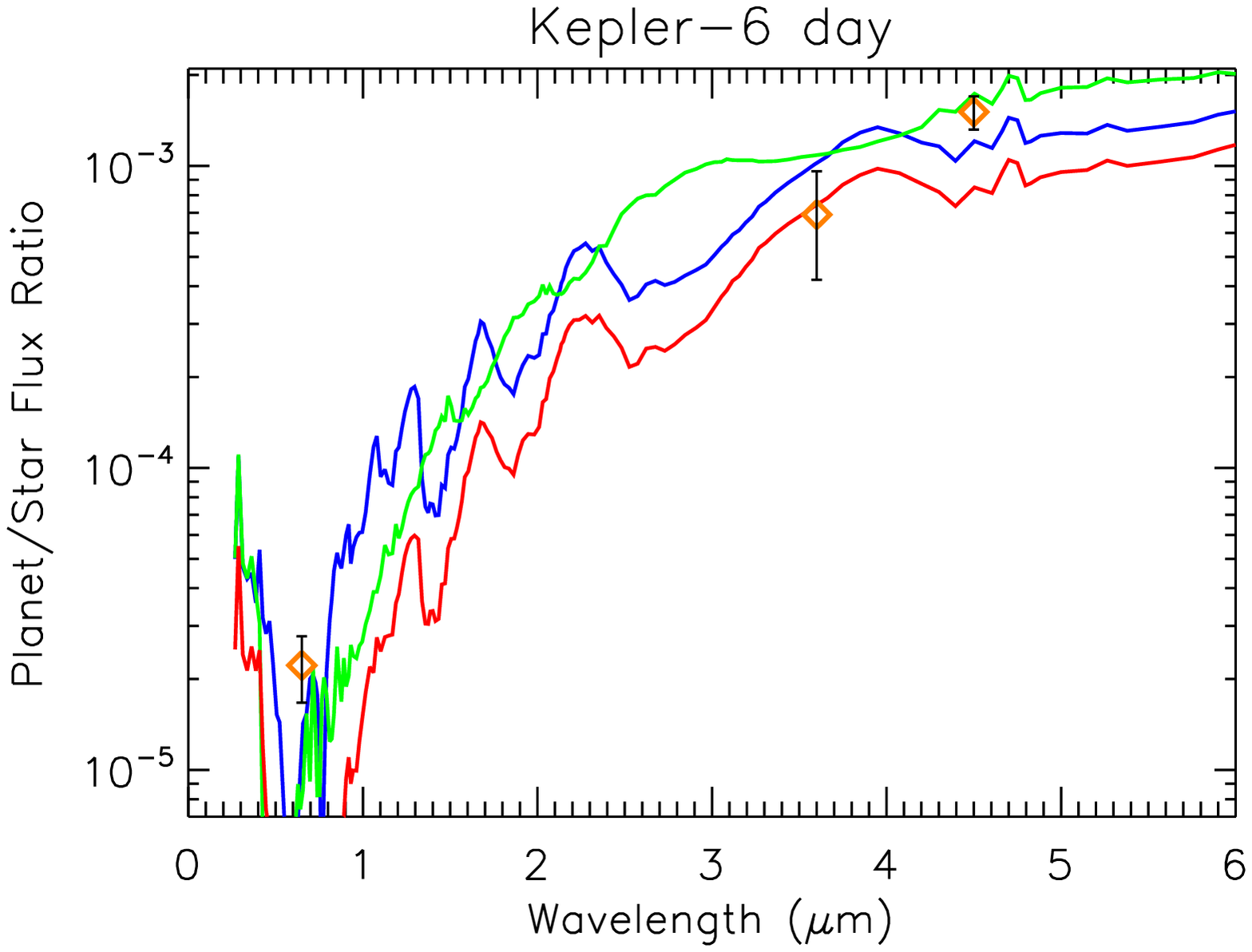}
 \caption{Day side planet-to-star flux ratios as function of the wavelength for three atmospheric models \citep{fortney08} per target. The orange diamonds and their error bars correspond to the \kepler\ and \spitzer\ observations. The blue models represent non-inverted atmospheres (no TiO) and no redistribution of the energy to the planetary night side. The green models represent inverted atmospheres (with TiO) and no redistribution of the energy to the planetary night side. The red models represent inverted atmospheres (with TiO) and with full redistribution of the energy to the planetary night side. Both targets are best fitted with non-inverted models which include dayside redistribution, with also an equivalent good fit for an inverted atmosphere for \ksb.}
   \label{fig:jonmodels}
\end{center}
\end{figure}

We compare the data to model predictions and select models with the best reduced \chisq. We note that this is not a fit involving adjustable parameters. 
All models for both targets show that the dayside is re-radiating the stellar flux efficiently, favoring low values for $f$.
Since our observations are weakly constraining, model comparisons for \ksb\ (Figure~\ref{fig:jonmodels}) suggest that both inverted and non-inverted atmospheric temperature profiles can reproduce the data. However this is not the case for \kcb\ where the models show no inversions.

For HJs such as \kc\ and \ks, both scattered stellar light and planetary thermal emission could contribute to the planet emergent flux in the {\it Kepler} bandpass \citep{lopez07}. The ratio of scattered starlight to thermal emission depends on the atmospheric composition. These objects could be too hot for condensates. Without a reflective condensate layer, photons in the \kepler\ bandpass are absorbed before being scattered. Therefore, the albedo of \kc\ and \ks\ are likely to be low, as observed here.


\cite{knutson10} shows that there could be a correlation between the host star activity level and the thermal inversion of the planetary atmosphere. Effectively, the strong XEUV irradiation from the active stellar host of HJs could deplete the atmosphere of chemical species responsible for producing inversions. The \caii\ line strengths are a good indicator of the stellar activity. We obtain Keck HIRES spectra for \kc\ and \ks, and estimate the line strengths of \sval=0.14\ and \logrhk=-4.97\ for \kc\ and \sval=0.16\ and \logrhk=-4.98\ for \ks. Both star are moderately quiet sub-giants, with \logrhk\ falling in-between cases for thermal inversion or no inversion (see discussion in \cite{knutson10}). We also evaluate the empirical index defined by \citet{knutson10} which provides a way to distinguish between the different hot-Jupiter atmospheres. \cite{knutson10} proposes an index value that could be correlated with the presence on a thermal inversion. Using the same definition, we find that the index = $-0.023 \pm 0.023$ and index = $0.040 \pm  0.033$ for \kc\ and \ks\ respectively. These values suggest that the atmosphere of \ksb\ is consistent with a weak thermal inversion and the one of \kcb\ with a non-inverted profile, which are both in agreement with the results of our present study.


%
As \kepler\ spacecraft continues to monitor the photometry of \kcb\ and \ksb, the future quarters of observations will be used to improve the signal-to-noise. This will provide better constraints on the occultations and may reveal the phase curves which would allow to derive a more accurate Bond albedo and thus provide better constrain the energy budget of these planetary atmospheres.


\begin{figure}[h!]
\begin{center}
 \includegraphics[width=3in]{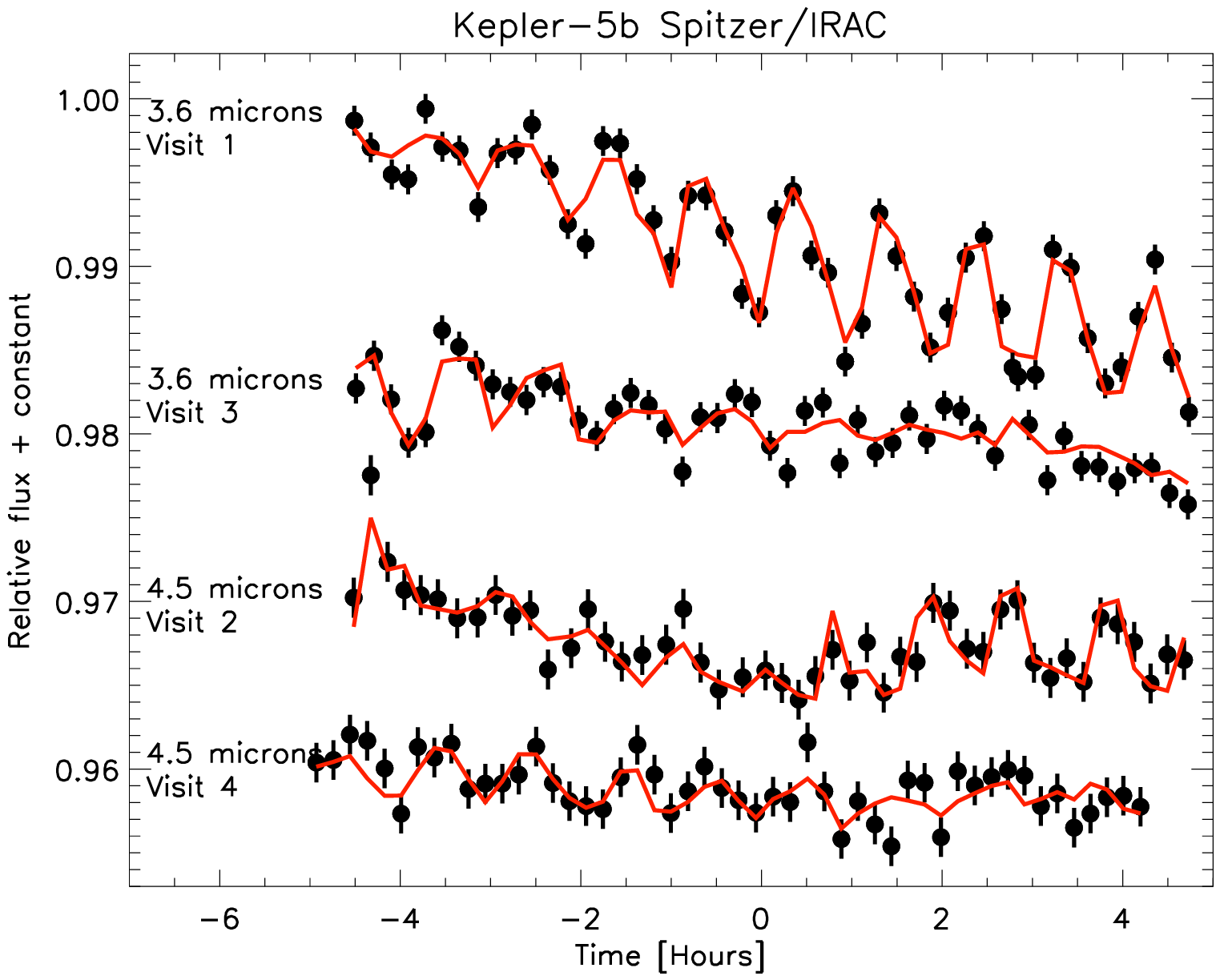}
 \includegraphics[width=3in]{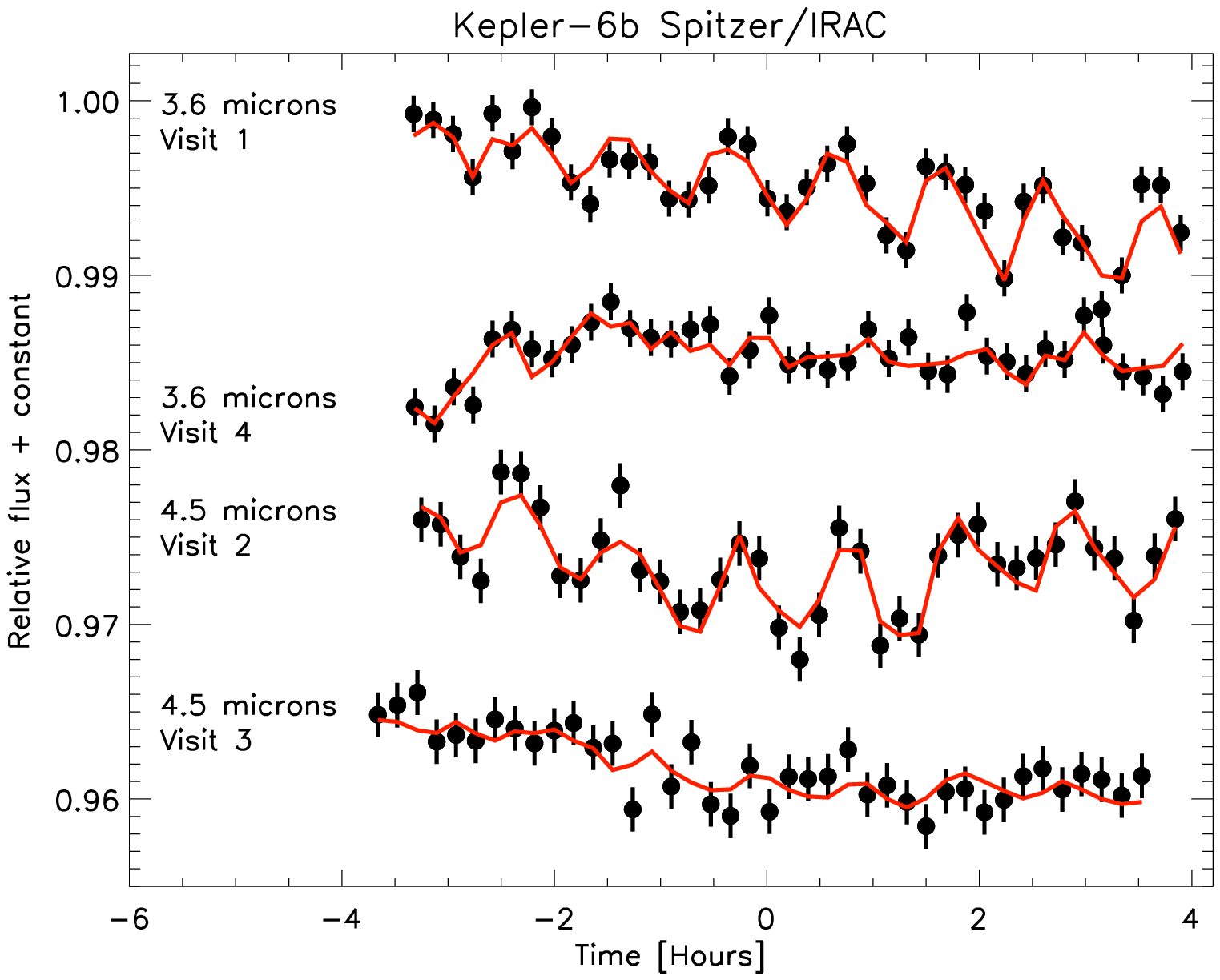}
 \caption{\spitzer\ raw occultation light-curves observed in the two currently available IRAC band-passes at 3.6 and 4.5~\micron, showing the two visits per bandpass (Table.~\ref{tab:spitzer_obs}). Plot for \kcb\ is on the left side and \ksb\ is on the right side. The data are binned in 12 minute intervals and offset in flux for display. The red solid line corresponds to the best fit models which include the time and position decorrelation as well as the model for the planetary occultation  (see details in Sect.~\ref{model_eclipse}).}
\label{fig:rawspitzerlightcurves}
\end{center}
\end{figure}

\begin{figure}[h!]
\begin{center}
\includegraphics[width=3in]{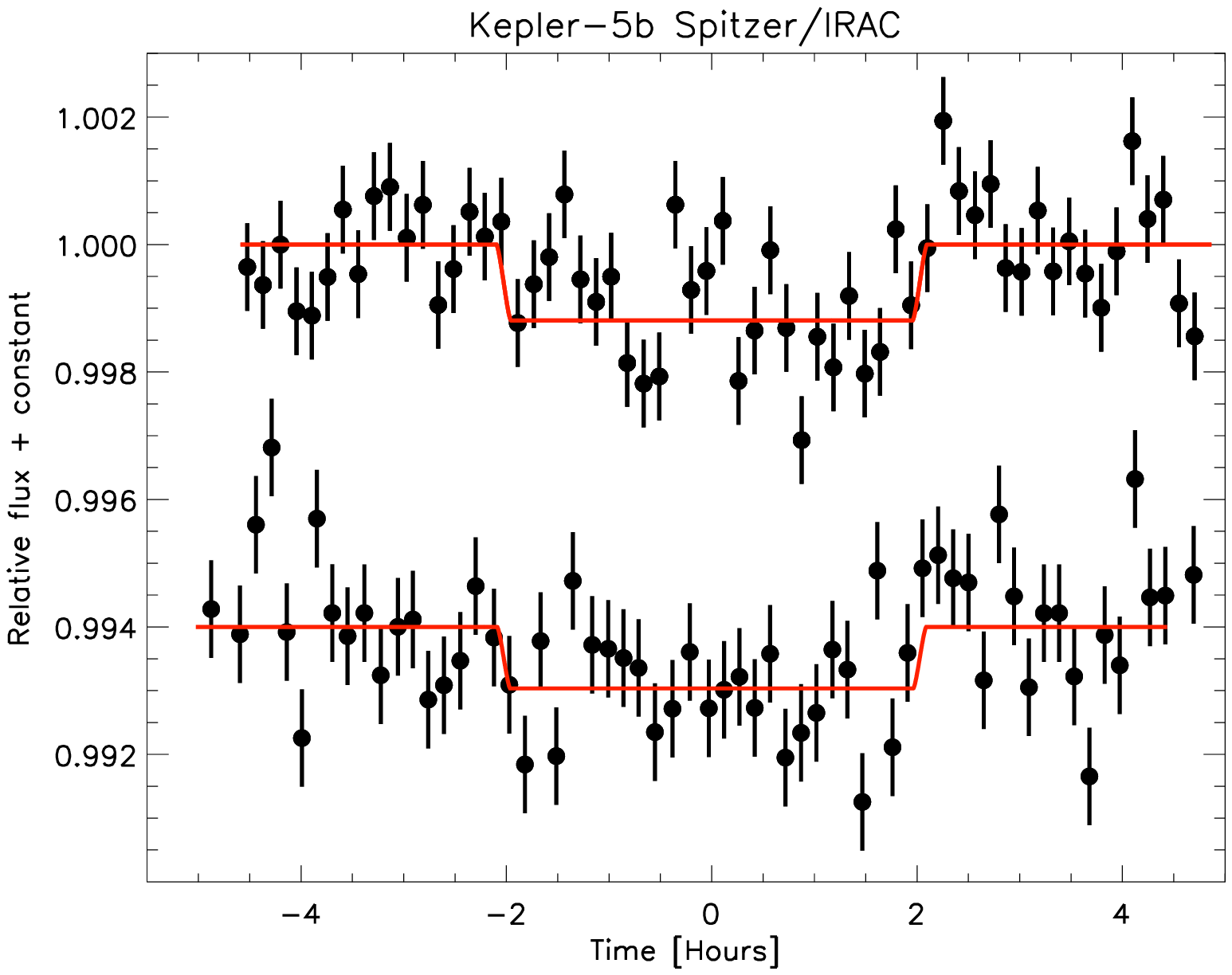}
\includegraphics[width=3in]{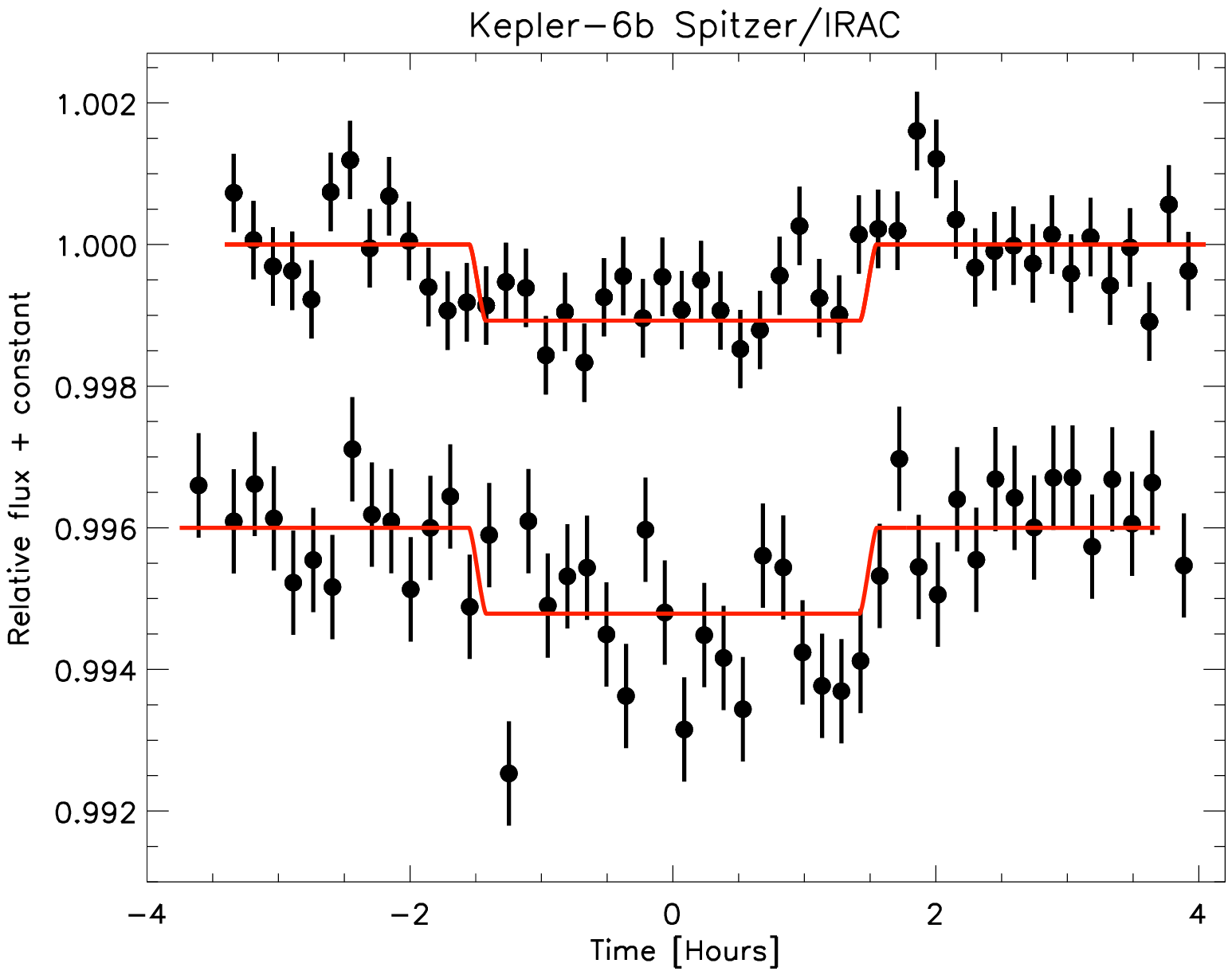}
 \caption{De-correlated, normalized  and combined per bandpass \spitzer\ occultation light-curves. The best fit are over-plotted in red solid line. The best fit depths with their associated errors are provided in Table.~\ref{tab:spitzer_obs}}
   \label{fig:spitzerlightcurves}
\end{center}
\end{figure}

\begin{figure}[h!]
\begin{center}
\includegraphics[width=3in]{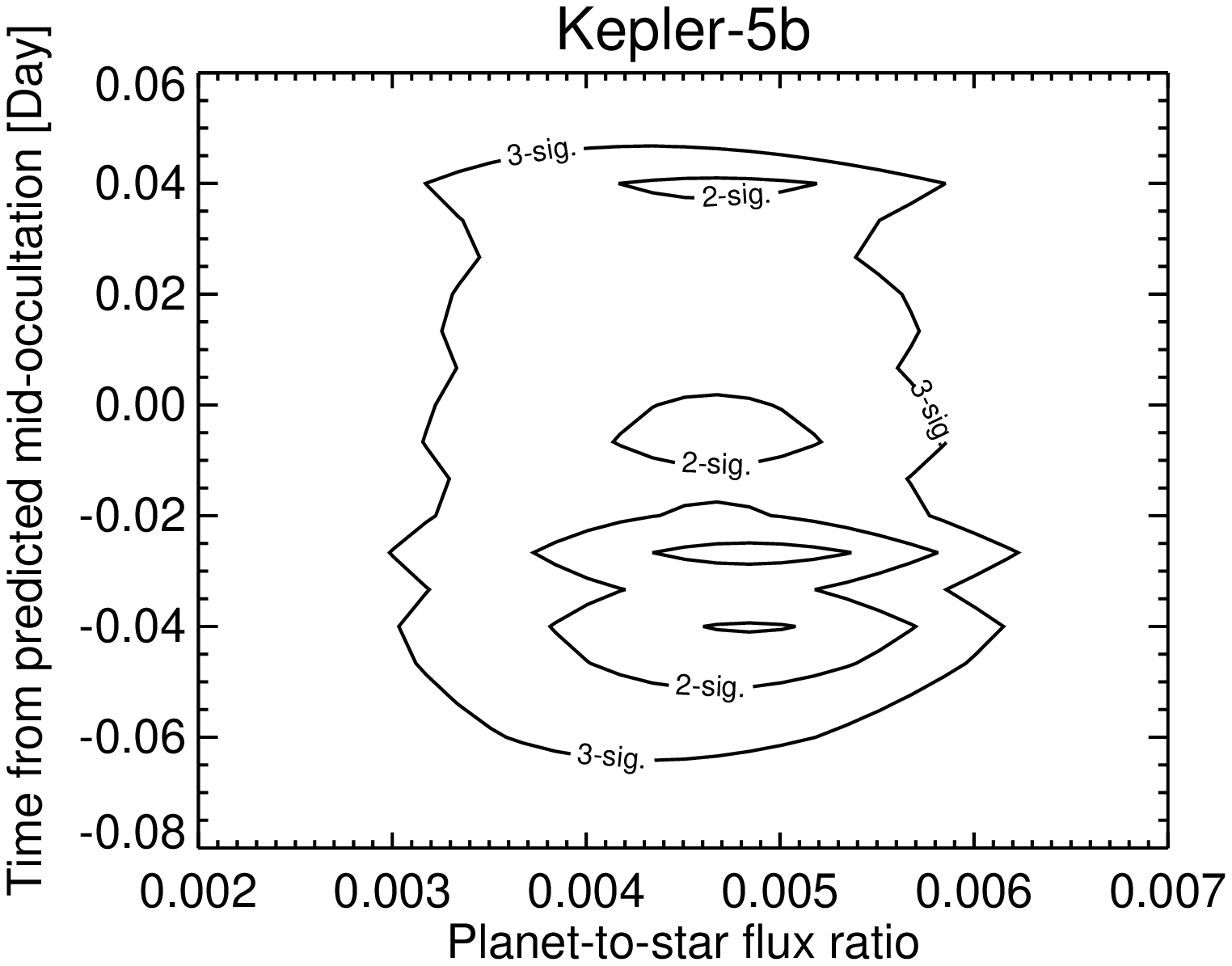}
\includegraphics[width=3in]{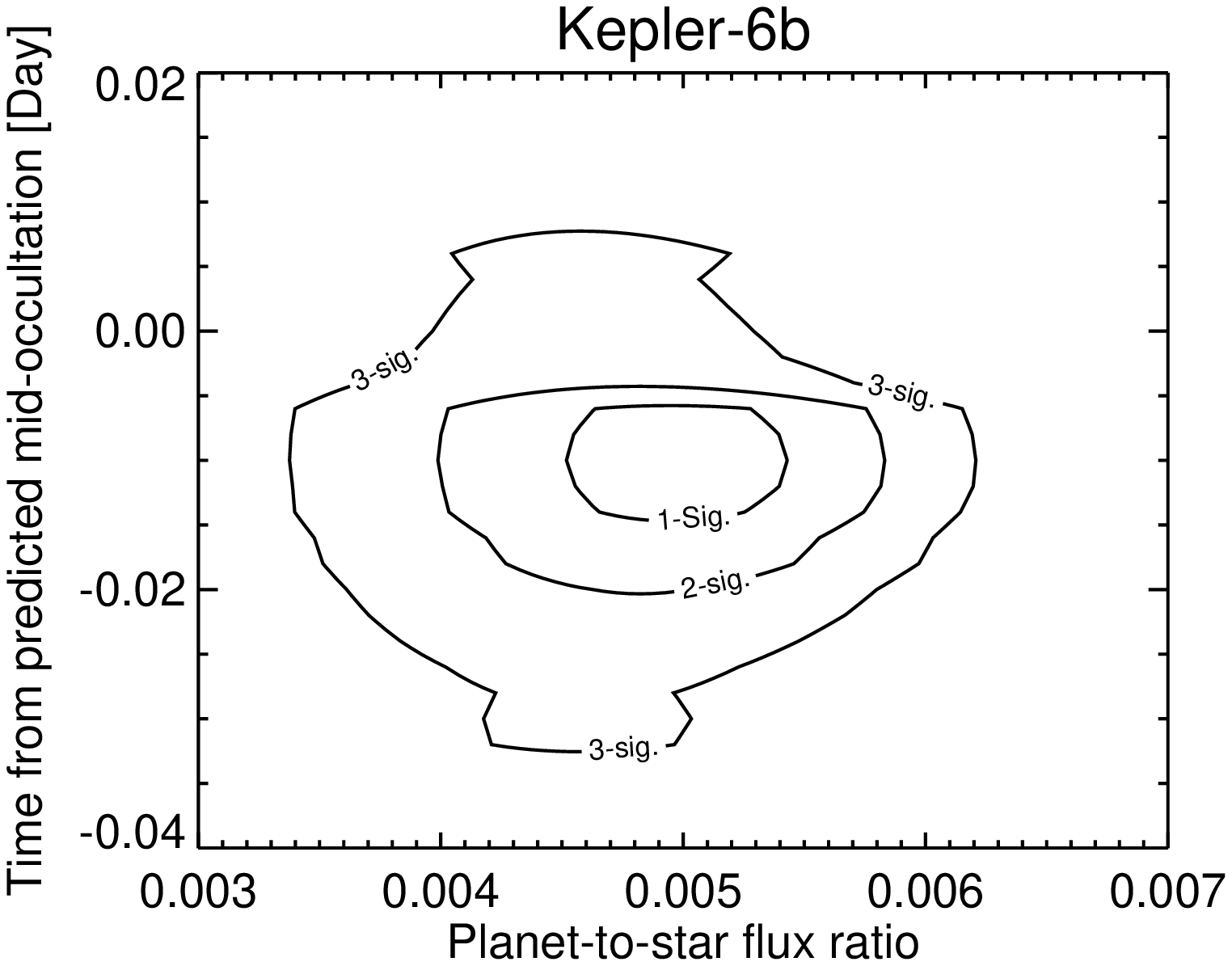}
 \caption{The \chisq\ spaces for mid-occultations time as function of the occultation depth, and the 1,2 and 3$\sigma$ confidence limits. Plot for \kcb\ is on the left side and \ksb\ is on the right side.}
   \label{fig:depthchisq}
\end{center}
\end{figure}

\begin{figure}[h!]
\begin{center}
  \includegraphics[width=5in]{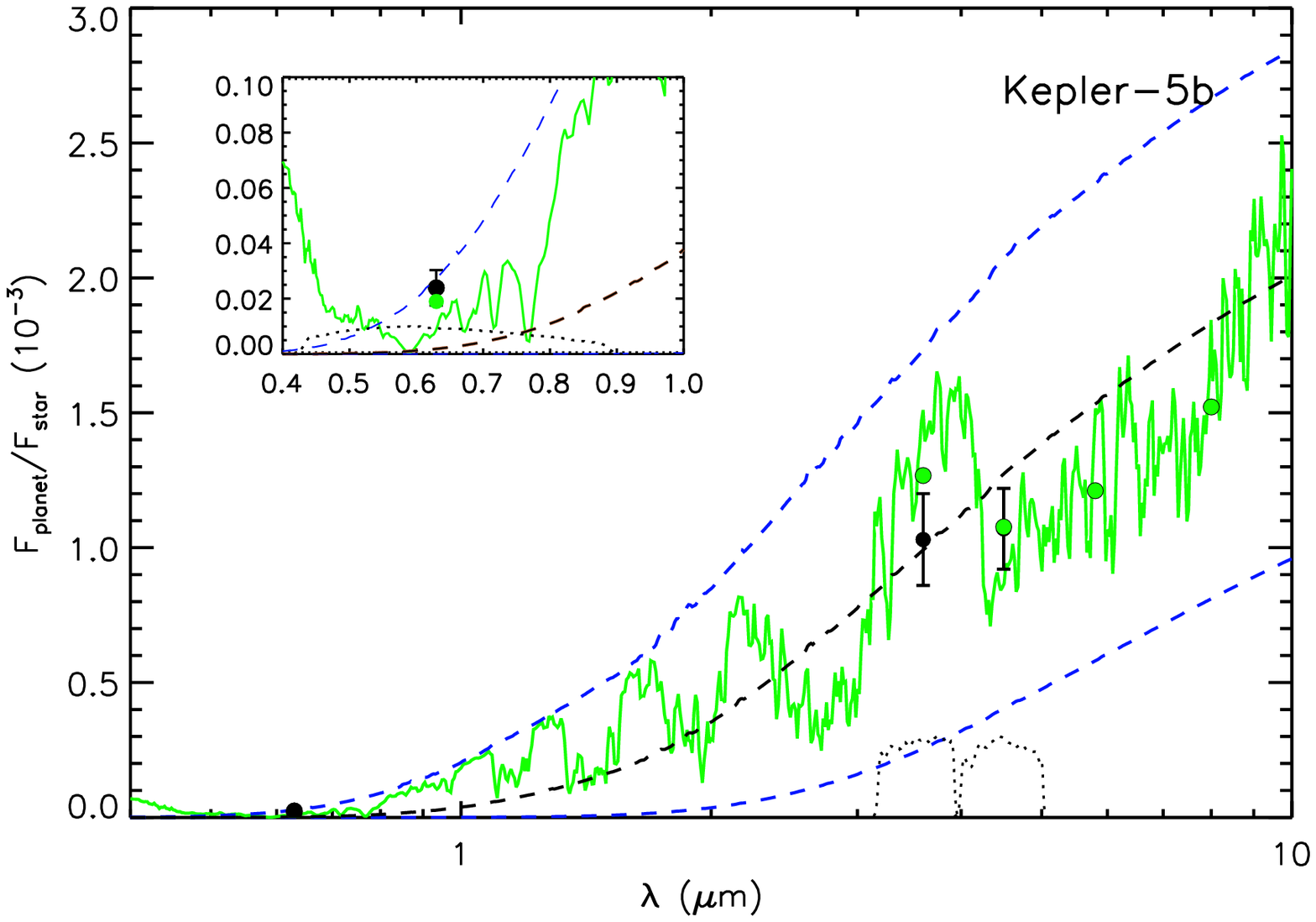}
 \includegraphics[width=5in]{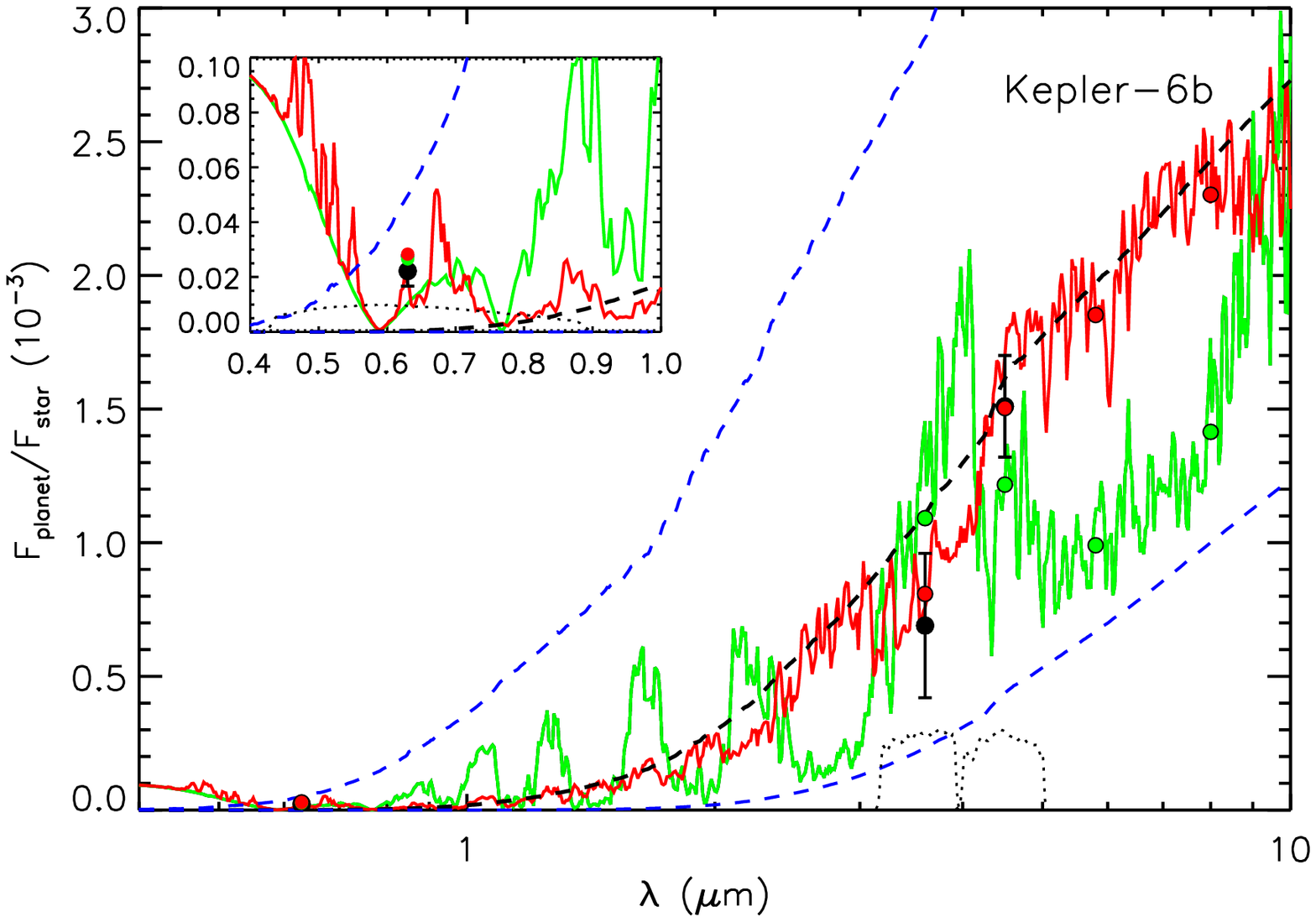}
 \caption{The \kepler\ and \spitzer\ occultations measured for \kcb\ and \ksb\ compared to hot-Jupiter atmospheric models \citep{madhu09}. The black points associated with their one-sigma error bars are the measurements. The solid lines correspond to the best fit models. The green and red points (without error bars) are the bandpass-integrated model values. The dashed line show black-body ratios for \spitzer\ measurements at various temperatures (see details in Sect.~\ref{atmo}). The insets are magnifications of the optical region of the spectra to reveal the \kepler\ measurements.}
   \label{fig:madhumodel}
\end{center}
\end{figure}

\begin{figure}[h!]
\begin{center}
   \includegraphics[width=3in]{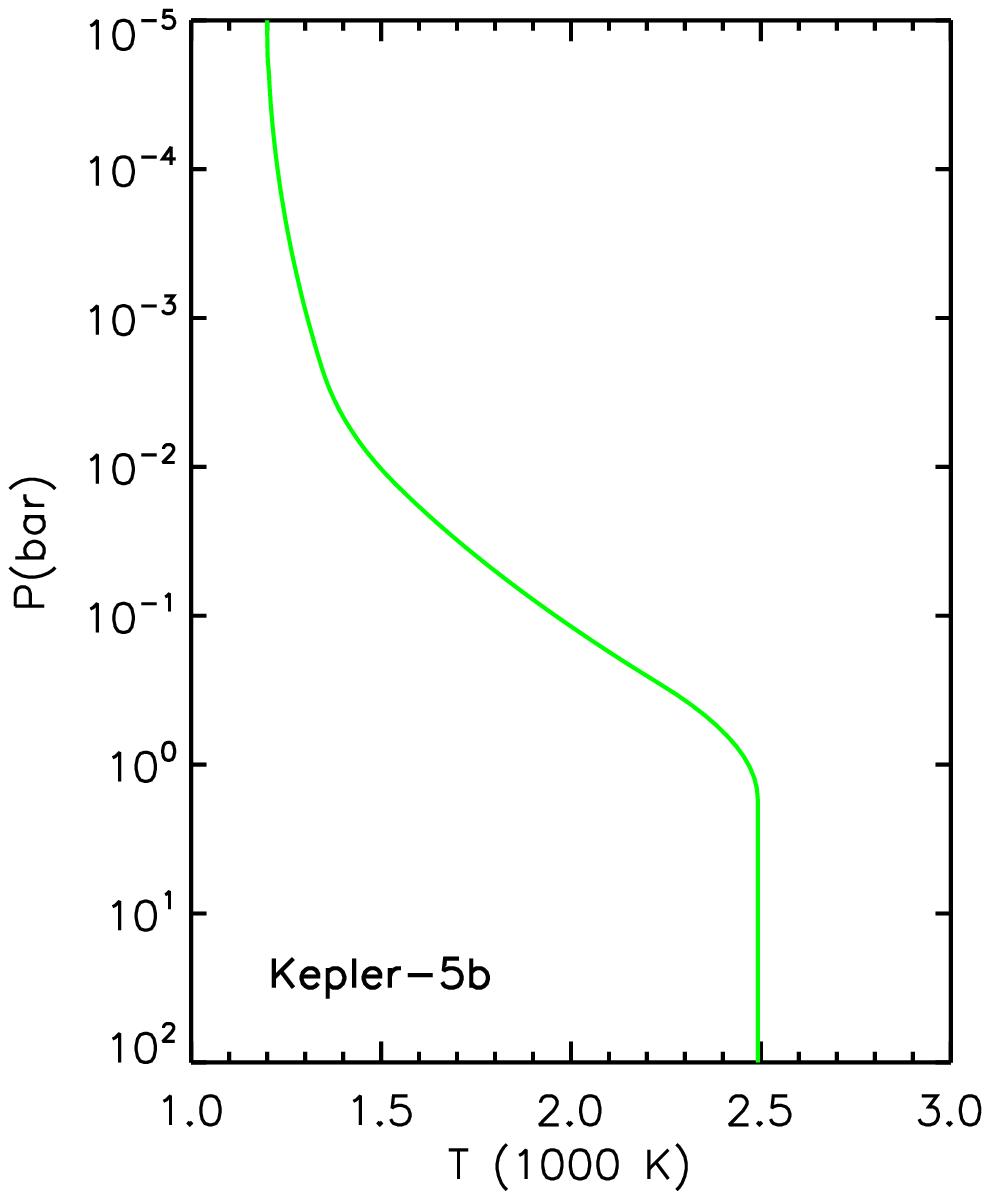}
     \includegraphics[width=3in]{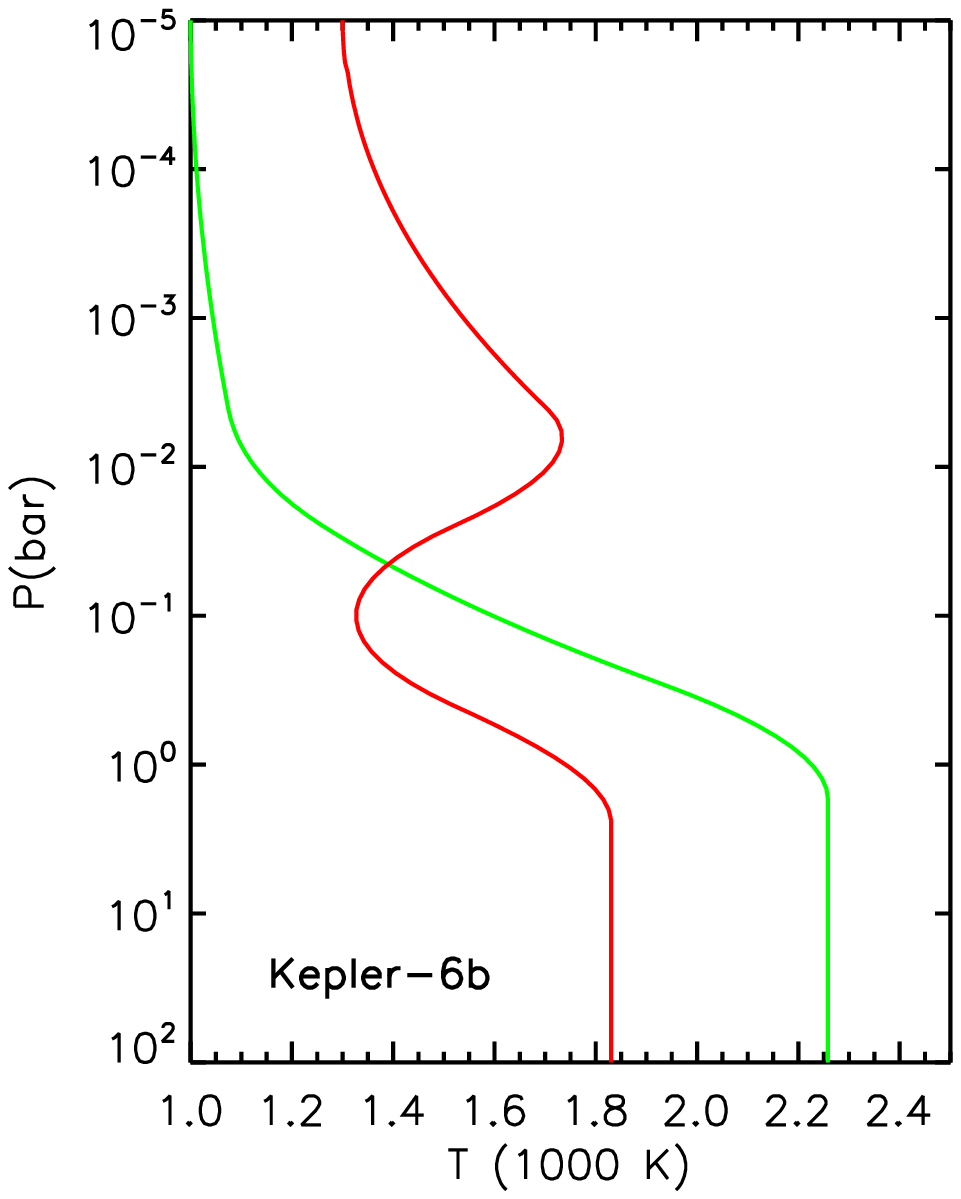}
 \caption{The temperature-pressure profiles for \kcb\ and \ksb\ corresponding to the model presented in Figure~\ref{fig:madhumodel}. The profile with a thermal inversion is over-plotted in red. Both profiles fits well \ksb\ observations.}
   \label{fig:madhumodel}
\end{center}
\end{figure}

\begin{deluxetable}{lcc}
\tabletypesize{\scriptsize}
\tablewidth{0pc}
\tablecaption{System Parameters for \kcb\ for \ksb\ obtained from the discovery papers. \label{tab:targets}}
\tablehead{\colhead{Parameter}	& 
\colhead{Value} }
\startdata
\sidehead{\em Kepler-5b from Koch et al.(2010)}   
     \\
Orbital period $P$ (d) & \ensuremath{3.548460\pm0.000032} \\ Mid-transit time $E$ (HJD) &
\ensuremath{2454955.90122\pm0.00021} \\ Scaled semi-major axis $a/\rstar$ & \ensuremath{6.06\pm0.14}
\\ Scaled planet radius \rpl/\rstar & \ensuremath{0.08195^{+0.00030}_{-0.00047}} \\ Impact
parameter $b \equiv a \cos{i}/\rstar$ & \ensuremath{0.393^{+0.051}_{-0.043}} \\ Orbital
inclination $i$ (deg) & \ensuremath{86\fdg3\pm0.5}  \\ Orbital eccentricity $e$ & $< 0.024$ \\
\sidehead{\em Stellar parameters}
Effective temperature \teff\ (K)                &\ensuremath{6297\pm60}         \\
Mass \mstar (\msun)                             & \ensuremath{1.374^{+0.040}_{-0.059}}       \\
Radius \rstar (\rsun)                           & \ensuremath{1.793^{+0.043}_{-0.062}}       \\
\sidehead{\em Planetary parameters}
Mass \mpl\ (\mjup)                              & \ensuremath{2.114^{+0.056}_{-0.059}}               \\
Radius \rpl\ (\rjup, equatorial)                & \ensuremath{1.431^{+0.041}_{-0.052}}                \\
Density \rhopl\ (\gcmc)                         & \ensuremath{0.894\pm0.079}               \\
Orbital semi-major axis $a$ (AU)                 & \ensuremath{0.05064\pm0.00070}           \\
Equilibrium temperature \teq\ (K)               & \ensuremath{1868\pm284}  \\ 

    \\
\hline              

\sidehead{\em Kepler-6b from Dunham et al.(2010)}       
\\
Orbital period $P$ (d)                          & \ensuremath{3.234723\pm0.000017}   \\
Mid-transit time $E$ (HJD)                       & \ensuremath{2454954.48636\pm0.00014} \\
Scaled semi-major axis $a/\rstar$                & \ensuremath{7.05^{+0.11}_{-0.06}}       \\
Scaled planet radius \rpl/\rstar                & \ensuremath{0.09829^{+0.00014}_{-0.00050}}   \\
Impact parameter $b \equiv a \cos{i}/\rstar$    & \ensuremath{0.398^{+0.020}_{-0.039}}    \\
Orbital inclination $i$                   & \ensuremath{86\fdg8\pm0.3}      \\
Orbital eccentricity $e$                        & 0 (adopted)       \\

\sidehead{\em Stellar parameters}
Effective temperature \teff\ (K)                & \ensuremath{5647\pm44}   \\
Mass \mstar (\msun)                             & \ensuremath{1.209^{+0.044}_{-0.038}} \\
Radius \rstar (\rsun)                           & \ensuremath{1.391^{+0.017}_{-0.034}}  \\
\sidehead{\em Planetary parameters}
Mass \mpl\ (\mjup)                              & \ensuremath{0.669^{+0.025}_{-0.030}}   \\
Radius \rpl\ (\rjup, equatorial)                            & \ensuremath{1.323^{+0.026}_{-0.029}}  \\
Density \rhopl\ (\gcmc)                         & \ensuremath{0.352^{+0.018}_{-0.022}} \\
Orbital semi-major axis $a$ (AU)                 & \ensuremath{0.04567^{+0.00055}_{-0.00046}}    \\
Equilibrium temperature \teq\ (K)               & \ensuremath{1500\pm200}
\enddata
\label{tab:stars}
\end{deluxetable}

\begin{deluxetable}{lccccccc}
\tabletypesize{\scriptsize}
\tablecaption{{\it Warm-Spitzer} observations. }
\tablewidth{0pt}
\tablehead{\colhead{Target} & \colhead{Visit} & \colhead{Wavelength} & \colhead{Obs. Date (UT)} & \colhead{Select. points} & \colhead{Depth} & \colhead{Weighted. Avg. depth} & \colhead{Bright. T (K)}}
\startdata
Kepler-5 & 1 & 3.6  &  2009-12-26  & 2455 & $0.091\pm0.022$\% &              -                   &        \\
Kepler-5 & 3 & 3.6 &  2010-06-16  & 2465 & $0.120\pm0.026$\% & $0.103\pm0.017$\% &   $1900\pm110$  \\
Kepler-5 & 2 & 4.5 &  2010-01-05  & 2540 & $0.102\pm0.023$\% &              -                   &      \\
Kepler-5 & 4 & 4.5 &  2010-07-12  & 2242 & $0.111\pm0.021$\% & $0.107\pm0.015$\% &  $1770\pm100$  \\
     \\
\hline              
     \\
Kepler-6 & 1 & 3.6  &  2009-12-30 & 2013 & $0.108\pm0.048$\% &              -                   &         \\
Kepler-6 & 4 & 3.6 &  2010-06-16 & 2012 & $0.051\pm0.032$\% & $0.069\pm0.027$\%  &  $1320\pm250$   \\ 
Kepler-6 & 2 & 4.5 &  2010-01-02 & 2022 & $0.123\pm0.027$\% &              -                    &     \\
Kepler-6 & 3 & 4.5 &  2010-01-13 & 1993 & $0.180\pm0.027$\% & $0.151\pm0.019$\%  &  $1700\pm120$  \\ \\
\enddata
\label{tab:spitzer_obs}
\end{deluxetable}

\begin{deluxetable}{lccccccc}
\tabletypesize{\scriptsize}
\tablecaption{{\it Kepler} observations. }
\tablewidth{0pt}
\tablehead{\colhead{Target} & \colhead{Quarters} & \colhead{Nbr of eclipses} & \colhead{Observing mode} & \colhead{$Fp/F_\star$ (ppm)}  & \colhead{Weighted. Avg. $Fp/F_\star$ (ppm)}  & \colhead{$A_{\rm g}$}}
\startdata
Kepler-5 & 0 to 1 & 11 &  Long cadence & $25 \pm 19$ &  & \\
Kepler-5 & 2 to 4 & 83 &  Short cadence & $21^{+5}_{-7}$  & $21 \pm 6$  & $0.12 \pm 0.04$ \\

     \\
\hline              
     \\
Kepler-6 & 0 to 1 & 12 &  Long cadence & $27 \pm 19$ &  &\\
Kepler-6 & 2 to 4 & 101 &  Short cadence & $21^{+8}_{-6}$   &  $22 \pm 7$ & $0.11 \pm 0.04$ \\

\enddata
\label{tab:kepler}
\end{deluxetable}

\acknowledgments

We would like to thank Jacob Bean and Brice-Olivier Demory for a variety of useful discussions.
This work is based on observations made with the Spitzer Space Telescope,
which is operated by the Jet Propulsion Laboratory, California Institute
of Technology under a contract with NASA. Support for this work was
provided by NASA through an award issued by JPL/Caltech. 
This work is also based on observations made with Kepler which was competitively selected as the tenth Discovery mission. Funding for this mission is provided by NASA's Science Mission Directorate. The authors would like to thank the many people who generously gave so much their time to make this Mission a success. Some of the data presented herein were obtained at the W.M. Keck Observatory, which is operated as a scientific partnership among the California Institute of Technology, the University of California and the National Aeronautics and Space Administration. The Observatory was made possible by the generous financial support of the W.M. Keck Foundation.

\end{document}